\documentclass{article}
\usepackage{comment}
\usepackage{geometry}
\usepackage{soul}
\usepackage{amssymb}
\usepackage{graphicx}
\usepackage{authblk}
\usepackage[utf8]{inputenc}
\usepackage{mathptmx}
\usepackage{amsmath}
\usepackage{epsfig,amsopn}
\usepackage{graphicx}
\usepackage{braket}
\usepackage{float}
\usepackage{enumerate}
\usepackage[normalem]{ulem}
\usepackage{todonotes}
\begin{document}

\title{Diversity of Sharp Restart}
\author{Iddo Eliazar\thanks{%
E-mail: \emph{eliazar@tauex.tau.ac.il}} \and Shlomi Reuveni\thanks{%
School of Chemistry, Tel-Aviv University, 6997801, Tel-Aviv, Israel.}\thanks{%
Center for the Physics and Chemistry of Living Systems. Tel Aviv University,
6997801, Tel Aviv, Israel.}\thanks{%
The Sackler Center for Computational Molecular and Materials Science, Tel
Aviv University, 6997801, Tel Aviv, Israel.}}
\maketitle

\begin{abstract}

When applied to a stochastic process of interest, a restart protocol alters the overall statistical distribution of the process' completion time; thus, the completion-time's mean and randomness change. The explicit effect of restart on the mean is well understood, and it is known that: from a mean perspective, deterministic restart protocols -- termed \emph{sharp restart} -- can out-perform any other restart protocol. However, little is known on the explicit effect of restart on randomness. This paper is the second in a duo exploring the effect of sharp restart on randomness: via a Boltzmann-Gibbs-Shannon entropy analysis in the first part, and via a diversity analysis in this part. Specifically, gauging randomness via diversity -- a measure that is intimately related to the Renyi entropy -- this paper establishes a set of universal criteria that determine: \textbf{A}) precisely when a sharp-restart protocol decreases/increases the diversity of completion times; \textbf{B}) the very existence of sharp-restart protocols that decrease/increase the diversity of completion times. Moreover, addressing jointly mean-behavior and randomness, this paper asserts and demonstrates when sharp restart has an aligned effect on the two (decreasing/increasing both), and when the effect is antithetical (decreasing one while increasing the other). The joint mean-diversity results require remarkably little information regarding the (original) statistical distributions of completion times, and are remarkably practical and easy to implement.

\bigskip\ 

\textbf{Keywords}: restart protocols; sharp restart; low-frequency and high-frequency resetting; diversity; Renyi entropy.

\bigskip\ 

\end{abstract}

\newpage

\section{\label{1}Introduction}

This paper is the second in a duo exploring the effect of \emph{restart protocols} on the inherent randomness of  task-completion durations of general stochastic processes. The first part of the duo addressed randomness via the perspective of the Boltzmann-Gibbs-Shannon entropy \cite{EntSR}. Elevating from this particular entropy to the general Renyi entropy, this part shall address randomness via the perspective of diversity (to be described below).

Restart research is a multidisciplinary scientific field which attracted significant interest in the recent years; a rather detailed literature survey of this field appears in the first part of the duo \cite{EntSR}. When a restart protocol is applied to a given stochastic process, it acts as follows: as long as the process does not accomplish its task, the protocol resets the process repeatedly. In order to time the resetting epochs the protocol uses a timer, which can be either stochastic or deterministic. Restart protocols with deterministic timers are termed, in short, \emph{sharp restart}.

Restart protocols have the potential of either expediting or impeding task-completion durations \cite{Pal2017}-\cite{Eva2020}. A main focus in restart research is thus set on mean-performance: determining when the application of restart protocols either decreases or increases mean task-completion durations \cite{Kus2014}-\cite{Yin2022}. Among general restart protocols, sharp restart assumes center stage due to the following key fact: from a mean-performance perspective, sharp restart can out-perform any other restart protocol \cite{Pal2017}-\cite{Che2018}. A comprehensive mean-performance analysis of sharp restart was established and presented in \cite{MP1SR}-\cite{MP2SR}. 

As stated and explained in the first part of this duo \cite{EntSR}, as well as in \cite{TBSR}, a mean-performance analysis addresses mean behavior alone, and it does not shed light on randomness. To visualize the situation, consider the iconic figure of statistics: the Gauss bell curve, which is quantified by two information items. The first item is the bell's center, which manifests the corresponding mean. The second item is the bell's width, which manifests the corresponding randomness. Knowing only the mean leaves us completely `in the dark' with regard to the inherent randomness of the bell-curve under consideration: we have no idea how peaked or spread-out the curve is about its center.

In essence, a similar situation holds with regard to sharp restart. Indeed, equipped only with a mean-performance analysis of sharp restart \cite{MP1SR}-\cite{MP2SR}, we are completely `in the dark' with regard to the effect of sharp restart on randomness. Namely, we do not know if the application of sharp restart decreases or increases the inherent randomness of task-completion durations. Enter this duo of papers, whose goal -- with regard to sharp restart -- is to bridge the knowledge gap between mean-performance and randomness. Attaining this goal will, in turn, lead to a combined mean-\&-randomness understanding of sharp restart.

Evidently, a precise quantification of randomness is a prerequisite in order to address the goal. There are several quantitative methods of gauging randomness, and an important such method -- which is used in various scientific fields -- is described as follows. Firstly, simulate two independent and identically distributed (IID) copies of the random quantity of interest (in our case, the task-completion duration of a given stochastic process). Secondly, calculate the `coincidence likelihood' of the two copies, i.e.: the likelihood that the two IID simulations yield the same outcome. Thirdly, set the reciprocal of the coincidence likelihood to be the quantitative gauge of randomness.

This gauge of randomness is known as Simpson's index in biology and ecology \cite{Sim}, as Hirschman's index in economics and competition law \cite{Hir1}, and as the participation ratio in physics \cite{BDH}-\cite{BD}. Also, in physics, the logarithm of the participation ratio is often referred to as the `collision entropy' \cite{BPP}. Notably, the participation ratio displays a set of properties (to be specified in the next section) that any reasonable measure of randomness should be expected to satisfy.

Underlying the participation ratio is a sample comprising two IID copies of the random quantity of interest. This raises the question: can the participation ratio be generalized so as to be based on a sample of arbitrary size? Namely, a sample comprising $\epsilon=2,3, \cdots$ copies of the random quantity of interest. The answer is affirmative, and the randomness gauge emanating from this generalization is termed \emph{diversity} \cite{Hil}-\cite{LL}. 

The diversity is a non-linear transformation of a `coincidence likelihood' that manifests the following scenario: the sample will appear as if it is deterministic -- with all the IID copies comprising the sample yielding the same outcome. Mathematically, the size $\epsilon$ of the sample can be further generalized: from the integer values $\epsilon=2,3, \cdots$ to the continuous values $\epsilon>1$. The logarithm of the diversity is the well-known \emph{Renyi entropy}, and the `size parameter' $\epsilon$ is the corresponding Renyi exponent \cite{Ren}-\cite{Zyc}. In the limit $\epsilon \rightarrow 1$, the Renyi entropy converges to the Boltzmann-Gibbs-Shannon entropy \cite{Bol}-\cite{Sha}.

As noted above, the first part of the duo presented a comprehensive Boltzmann-Gibbs-Shannon entropy analysis of sharp restart \cite{EntSR}. This part continues with a comprehensive diversity analysis of sharp restart. Specifically, this paper shall establish a set of closed-form results that address sharp restart with: general timers; `fast timers' that manifest high-frequency resetting; and `slow timers' that manifest low-frequency resetting -- a regime that attracted much scientific attention \cite{Slow_Time_1}-\cite{Slow_Time_11}. Moreover, this part shall further establish a set of closed-form results that determine the very existence of timers with which sharp restart either decreases or increases the diversity of task-completion durations.

A highlight of this paper is a set of joint mean-diversity results regarding sharp restart with high-frequency and low-frequency resetting. The results assert that sharp restart can have a similar effect, as well as an opposite effect, on the mean and the diversity. Specifically, there are scenarios in which sharp restart decreases/increases both the mean and the diversity. And, there are scenarios in which sharp restart decreases the mean while increasing the diversity, and vice-versa. The mean-diversity results characterize all four mean-diversity scenarios via universal criteria that: require very little statistical information, are remarkably simple and transparent, and are remarkably practical and easy-to-implement. 

This paper is organized as follows. Section \ref{2} sets the stage: it recaps sharp restart, and reviews the notion of diversity. Section \ref{3} presents a general statistical analysis of the effect of sharp restart on diversity. The statistical analysis determines, for any given timer, when sharp restart decreases/increases diversity; also, the statistical analysis characterizes the case in which sharp restart neither decreases nor increases diversity. Section \ref{4} establishes results that determine the very existence of timers with which sharp restart decreases/increases diversity. Section \ref{5} -- using the reliability-engineering notion of hazard rate -- presents an asymptotic analysis of the effect of sharp restart on diversity. The asymptotic analysis addresses two cases: high-frequency resetting, which uses very small timers; and low-frequency resetting, which uses very large timers. For such timers, the asymptotic analysis determines when sharp restart decreases/increases diversity. Combining the diversity results of section \ref{5} together with corresponding mean-performance results established in \cite{MP1SR}, section \ref{6} characterizes -- for high-frequency and low-frequency resetting -- the joint effect of sharp restart on the mean and the diversity of task completion times. Section \ref{7} concludes with a summary of the key results presented along the paper. The results' derivations are detailed in the Methods.

\section{\label{2}Setting the stage}

This section sets the stage for a diversity analysis of sharp restart. Following \cite{MP1SR}-\cite{TBSR}, subsection \ref{21} tersely recaps sharp restart. Subsection \ref{22} reviews the notion of diversity \cite{Hil}-\cite{LL}. And, just before embarking on the diversity analysis, subsection \ref{23} explains the `raison d'être' of this analysis.

\subsection{\label{21}Sharp restart}

\emph{Sharp restart} is an \emph{algorithm} that is described as follows \cite{MP1SR}-\cite{TBSR}. There is a general task with completion time $T$, a positive-valued random variable. To this task a three-steps algorithm, with a positive deterministic timer $\tau $, is applied. Step I: initiate simultaneously the task and the timer. Step II: if the task is accomplished up to the timer's expiration -- i.e. if $T\leq \tau $ -- then stop upon completion. Step III: if the task is not accomplished up to the timer's expiration -- i.e. if $T>\tau $ -- then, as the timer expires, go back to Step I.

The sharp-restart algorithm generates an iterative process of independent and statistically identical task-completion trials. This process halts during its first successful trial, and we denote by $T_{R}$ its halting time. Namely, $T_{R}$ is the overall time it takes -- when the sharp-restart algorithm is applied -- to complete the task. The sharp-restart algorithm is a \emph{non-linear mapping} whose input is the random variable $T$, whose output is the random variable $T_{R}$, and whose parameter is the deterministic timer $\tau $.

Henceforth, we set the task-completion process to start at time $t=0$; thus, the process takes place over the non-negative time axis $t\geq 0$. Along this paper we use the following notation regarding the input's statistics: distribution function, $F\left( t\right) =\Pr \left( T\leq t\right) $ ($t\geq 0$); survival function, $\bar{F}\left( t\right) =\Pr \left( T>t\right)$ ($t\geq 0$); and density function, $f\left( t\right) =F^{\prime }\left(t\right) =-\bar{F}^{\prime }\left( t\right) $ ($t>0$). The input's density function is considered to be positive-valued over the positive half-line: $f\left( t\right) >0$ for all $t>0$.

In terms of the input's survival and density functions, the output's density function $f_{R}\left( t\right) $ admits the following representation \cite{TBSR}:
\begin{equation}
f_{R}\left( \tau n+u\right) =\bar{F}\left( \tau \right) ^{n}f\left( u\right), \label{211}
\end{equation}
where $n=0,1,2,\cdots $, and where $0\leq u<\tau $. Indeed, in order that the output $T_{R}$ be realized at time $t=\tau n+u$ we need that: \textbf{A}) the first $n$ task-completion trials be unsuccessful; and \textbf{B}) the task be accomplished right at the time epoch $u$ of the $(n+1)^{\text{th}}$ task-completion trial. Event \textbf{A} occurs with probability $\bar{F}\left( \tau\right) ^{n}$, event \textbf{B} occurs with likelihood $f\left( u\right) $, and hence: the likelihood that the output be realized right at time $t=\tau n+u$ is given by the right-hand side of Eq. (\ref{211}).

Due to the structure of the sharp-restart algorithm, its output admits the stochastic representation
\begin{equation}
T_{R}=\tau N+U\text{ ,}  \label{212}
\end{equation}
where: $N=\lfloor T_{R}/\tau \rfloor $ is the number of unsuccessful task-completion trials; and $U=T_{R}-\tau N$ is the time epoch, within the successful trial, at which the task is accomplished. The random variables $N$ and $U$ are mutually independent, and their statistics are as follows \cite{TBSR}. The random variable $N$ is Geometrically-distributed with success probability $p=F\left( \tau \right) $, i.e.: $\Pr \left(N=n\right) =\left( 1-p\right) ^{n}p$ ($n=0,1,2,\cdots $). The random variable $U$ is equal, in law, to the input $T$ -- given the information that the input is no-larger than the timer, $\left\{ T\leq \tau \right\} $; consequently, the statistical distribution of the random variable $U$ is governed by the density function $f\left( u\right) /F\left( \tau \right) $ ($0<u<\tau $).

\subsection{\label{22}Diversity}

Perhaps the most common quantitative gauge of randomness is the standard deviation. Specifically, the standard deviation $\sigma \left[ X\right] $ of a real-valued random variable $X$ is the square root of the random variable's variance. A principal property of the standard deviation is its response to affine transformations: $\sigma \left[ aX+b\right] =\left\vert a\right\vert \cdot \sigma \left[ X\right] $, where $a$
and $b$ are real parameters. 

The aforementioned principal property captures four different features that one would reasonably expect from any non-negative gauge of randomness (of real-valued random variables). \textbf{I}) The randomness of a deterministic, i.e. constant, random variable is zero. \textbf{II}) Translating a random variable does not change its randomness: $\sigma[X+b]=\sigma[X]$, where $b$ is a real translation parameter. \textbf{III}) Mirroring a random variable does not change its randomness: $\sigma[-X]=\sigma[X]$. \textbf{IV}) Changing the scale of a random variable changes its randomness by the very same scale: $\sigma[s \cdot X]=s \cdot \sigma[X]$, where $s$ is a positive scale parameter.

The standard deviation is constructed from a Euclidean-geometry perspective. Shifting the perspective from geometric to probabilistic -- while maintaining the aforementioned principal property in mind -- we shall now show how to construct a gauge of randomness that quantifies the inherent `determinism'/`stochasticity' of random variables.

Consider a real-valued random variable $X$, whose statistical distribution is governed by the density function $\varphi \left( x\right) $ ($-\infty<x<\infty $). Further consider a sample $\left\{ X_{1},\cdots ,X_{\epsilon}\right\} $ comprising $\epsilon $ IID copies of the random variable $X$, where $\epsilon $ is an integer that is greater than one ($\epsilon=2,3,\cdots $). The likelihood of the coincidence event $X_{1}=\cdots=X_{\epsilon }$ is: 
\begin{equation}
\mathcal{C}_{\epsilon }\left[ X\right] =\int_{-\infty }^{\infty}\varphi \left( x\right) ^{\epsilon }dx.  \label{221}
\end{equation}

The term $\mathcal{C}_{\epsilon }\left[ X\right] $ appearing in Eq. (\ref{221}) is a \emph{coincidence likelihood} that can be interpreted as follows: it is the likelihood that the sample $\left\{ X_{1},\cdots ,X_{\epsilon }\right\} $ will appear \emph{as if} it was generated deterministically. Hence, the coincidence likelihood $\mathcal{C}_{\epsilon }\left[ X\right] $ is a measure of the inherent `determinism' of the random variable $X$. In turn, the reciprocal of the coincidence likelihood, $1/\mathcal{C}_{\epsilon }\left[ X\right] $, is a measure of the inherent `stochasticity' of the random variable $X$.

Now, consider the aforementioned affine transformation, $X\mapsto aX+b$, where $a$ and $b$ are real parameters. It is straightforward to check that the response of the randomness measure $1/\mathcal{C}_{\epsilon }\left[ X\right]$ to the affine transformation is: $1/\mathcal{C}_{\epsilon }\left[aX+b\right] =\left\vert a\right\vert ^{\epsilon -1}/\mathcal{C}_{\epsilon }\left[ X\right] $. Consequently, if we want to obtain a response (to the affine transformation) that is identical to that of the standard deviation -- we need to raise the randomness measure $1/\mathcal{C}_{\epsilon }\left[ X\right] $ by the power $1/(\epsilon -1)$. Doing so we arrive at the \emph{diversity} \cite{Hil}-\cite{LL}: 
\begin{equation}
\mathcal{D}_{\epsilon }\left[ X\right] =\left\{ \frac{1}{\mathcal{C}_{\epsilon }\left[ X\right] }\right\} ^{\frac{1}{\epsilon -1}}=\left[\int_{-\infty }^{\infty }\varphi \left( x\right) ^{\epsilon }dx\right]^{1/(1-\epsilon )}.  \label{222}
\end{equation}

The diversity $\mathcal{D}_{\epsilon }\left[ X\right] $ is a measure of the inherent `stochasticity' of the random variable $X$. The diversity $\mathcal{D}_{\epsilon }\left[ X\right] $ exhibits the principal property that the standard deviation exhibits -- the following response to affine transformations: $\mathcal{D}_{\epsilon }\left[ aX+b\right] =\left\vert a\right\vert \cdot \mathcal{D}_{\epsilon }\left[ X\right] $, where $a$ and $b$ are real parameters. 

In the special case $\epsilon =2$ -- which manifests a sample comprising two IID copies of the random variable $X$ -- the diversity is the reciprocal of the coincidence likelihood: $\mathcal{D}_{2}\left[ X\right]=1/\mathcal{C}_{2}\left[ X\right]$. This diversity is known as: \emph{Simpson's index} \cite{Sim} in biology and ecology -- where it is used as a measure of biodiversity; \emph{Hirschman's index} \cite{Hir1} in economics and competition law -- where it is applied as a measure of concentration;\footnote{Hirschman's index was reinvented by Herfindahl, and is commonly -- yet
mistakenly -- referred to as the Herfindahl index or the Herfindahl-Hirschman index \cite{Hir2}.} and \emph{participation ratio} \cite{BDH}-\cite{BD} in physics -- where it is used as a measure of localization. Also, this diversity has a profound relation to the overall correlation of moving-average random processes \cite{MAAD}.

Observing Eq. (\ref{222}) it is evident that the inverse relationship between the diversity $\mathcal{D}_{\epsilon }\left[ X\right] $ and the coincidence likelihood $\mathcal{C}_{\epsilon }\left[ X\right] $ holds for all $\epsilon >1$. Thus, mathematically, we can extend Eq. (\ref{222}) from the integer range $\epsilon =2,3,\cdots $ to the continuous range $\epsilon >1$. Doing so, the logarithm of the diversity $\ln ( \mathcal{D}_{\epsilon} \left[ X\right] ) $ is the \emph{Renyi entropy} of the random variable $X$ \cite{Ren}-\cite{Zyc}, and $\epsilon$ is the corresponding Renyi exponent. In the special case $\epsilon =2$ -- which manifests a sample of size two -- the logarithm of the diversity, $\ln ( \mathcal{D}_{2} \left[ X\right] )= - \ln ( \mathcal{C}_{2} \left[ X\right] )$, is the \emph{collision entropy} of the random variable $X$ \cite{BPP}.

Yet another special case of the Renyi entropy is obtained in the Renyi-exponent limit $\epsilon \rightarrow 1$, which manifests a sample of size one. Indeed, in this limit the diversity $\mathcal{D}_{\epsilon }\left[ X\right] $ converges to the \emph{perplexity} of the random variable $X$ \cite{JMB}. And, the Renyi entropy $\ln ( \mathcal{D}_{\epsilon }\left[ X\right] ) $ converges to the \emph{Boltzmann-Gibbs-Shannon entropy} of the random variable $X$ \cite{Bol}-\cite{Sha}.

The first part of this duo \cite{EntSR} explored the effect of the sharp-restart algorithm on randomness via the perspective of the Boltzmann-Gibbs-Shannon entropy. Elevating from this particular entropy to the general Renyi entropy, this part shall explore the effect of the sharp-restart algorithm on randomness via the perspective of diversity.

\subsection{\label{23}Raison d'être}

With Eq. (\ref{211}) at hand, the statistical analysis of sharp restart is -- in theory -- complete. Indeed, Eq. (\ref{211}) presents an explicit closed-form formula for the output's density function $f_{R}(t)$ in terms of: the input's density and survival functions, $f(t)$ and $\bar{F}(t)$; and the sharp-restart timer parameter $\tau$. Thus, given the input's statistical distribution, and setting the timer parameter, one can calculate the output's density function. In turn, various statistics of the output can be further calculated, e.g.: the output's mean (which was the focus of the studies \cite{MP1SR}-\cite{MP2SR}), as well as higher-order moments of the output; the output's Boltzmann-Gibbs-Shannon entropy (which was the focus of the first part of this duo \cite{EntSR}); and the output's diversity (which is the focus of this paper). So, why bother with the mean-performance analyses of \cite{MP1SR}-\cite{MP2SR}, and with the randomness analyses of this duo?

Well, as stated in the opening sentence of this subsection, the aforementioned `calculational approach' is fine \emph{in theory}, yet in practice this approach is rather impractical. Firstly, consider the ideal situation in which the input's statistical distribution is known in full detail. In this situation the calculation of Eq. (\ref{211}) -- which is performed case-by-case, per a given input and per a given timer -- provides no general insights regarding the effect of sharp restart on mean-performance and on randomness. Secondly, consider situations in which full information regarding the input's statistical distribution is lacking. In `real life' the latter situations are prevalent, and in such situations Eq. (\ref{211}) is outright unusable. Thus, Eq. (\ref{211}) is a foundational theoretical result, but it is not a practical `working tool'.

As established in the mean-performance analyses of \cite{MP1SR}-\cite{MP2SR} and in the entropy analysis of \cite{EntSR}, and as shall be established here: `standing on the shoulders' of Eq. (\ref{211}), it is possible to construct highly insightful and highly implementable results that altogether circumvent the need to use Eq. (\ref{211}) directly. Indeed, with remarkably little information about the input's statistical distribution -- rather than full knowledge -- explicit criteria can be drawn regarding the effect of sharp restart on mean-performance, on randomness, and \emph{jointly} on mean-performance and randomness. Moreover, these explicit criteria are universal: they are applicable to \emph{any} input, with \emph{any} statistical distribution.

\section{\label{3} Diversity analysis}

Having recapped the sharp-restart algorithm, and having reviewed the notion of diversity, we are all set to explore the diversity of sharp restart. This section shall present a statistical analysis of how the application of the sharp-restart algorithm changes the diversity of a given input to that of its output.

In terms of the input's survival and density functions, the output's
coincidence likelihood is: 
\begin{equation}
\mathcal{C}_{\epsilon }\left[ T_{R}\right] =\frac{1}{1-\bar{F}\left( \tau \right) ^{\epsilon }}\int_{0}^{\tau }f\left( u\right) ^{\epsilon }du.
\label{301}
\end{equation}
Eq. (\ref{301}) is obtained by applying the coincidence formula of Eq. (\ref {221}) to the density formula of Eq. (\ref{211}). The derivation of Eq. (\ref{301}) is detailed in the Methods. Note that setting $\tau=\infty$ in Eq. (\ref{301}) yields the input's coincidence likelihood, $\mathcal{C}_{\epsilon }[T]=\int_{0}^{\infty }f\left( u\right) ^{\epsilon }du$.

In the Methods it is further shown that Eq. (\ref{301}) can be re-written as the product of the coincidence likelihoods of the random variables $N$ and $U$ of Eq. (\ref{212}), i.e.: $\mathcal{C}_{\epsilon }\left[ T_{R}\right] =\mathcal{C}_{\epsilon }\left[ N \right] \cdot \mathcal{C}_{\epsilon }\left[ U\right]$. Consequently, the product-form of the coincidence likelihoods carries on to the corresponding diversities: 
\begin{equation}
\mathcal{D}_{\epsilon }\left[ T_{R}\right] =\mathcal{D}_{\epsilon }\left[ N \right] \cdot \mathcal{D}_{\epsilon }\left[ U\right].  \label{302}
\end{equation}

To appreciate Eq. (\ref{302}) we emphasize that, in general, Eq. (\ref{212}) need \emph{not} imply Eq. (\ref{302}). Namely, the diversity of a linear combination of (two) independent random variables is \emph{not} equal, in general, to the product of the diversities of the (two) random variables. Here, Eq. (\ref{212}) implies Eq. (\ref{302}) due to the very specific structure of the sharp-restart algorithm: the one-to-one correspondence between the output $T_{R}$ and the pair $(N,U)$, and the fact that the random variables $N$ and $U$ are independent random variables.\footnote{Indeed, the independence fact implies that the diversity of the pair $(N,U)$ is the product of the diversities of $N$ and $U$. And, in turn, the one-to-one correspondence implies that the diversity of the output $T_{R}$ is the diversity of the pair $(N,U)$.}

In what follows we denote by $\gamma =\mathcal{C}_{\epsilon }\left[ T\right] $ the input's coincidence likelihood, and by $\delta =\mathcal{D}_{\epsilon } \left[ T\right] $ the input's diversity. Also, we denote by $C\left( \tau \right) =\mathcal{C}_{\epsilon }\left[ T_{R}\right] $ the output's coincidence likelihood, and by $D\left( \tau \right) =\mathcal{D}_{\epsilon }\left[ T_{R}\right] $ the output's diversity; this notation underscores the fact that the output's coincidence likelihood and diversity are functions of the timer $\tau $, the parameter of the sharp-restart algorithm. The goal of this study is to explore the effect of the sharp-restart algorithm on diversity. To that end we use the following terminology:

\begin{enumerate}
\item[$\bullet $] Sharp restart with timer $\tau $ \emph{decreases diversity} if the output's diversity is smaller than the input's
diversity, $D\left( \tau \right) <\delta $. Due to the inverse relation between diversity and  coincidence, this is equivalent to the output's coincidence likelihood being larger than that of the input, $C\left(\tau \right) >\gamma $.

\item[$\bullet $] Sharp restart with timer $\tau $ \emph{increases diversity} if the output's diversity is larger than the input's diversity, $D\left( \tau \right) >\delta $. Due to the inverse relation between diversity and  coincidence, this is equivalent to the output's coincidence likelihood being smaller than that of the input, $C\left(\tau \right) <\gamma $.
\end{enumerate}

As an illustrative demonstration of a diversity analysis, consider an input whose distribution is Pareto type II. Pareto distributions \cite{Par}, which comprise four types, are the principal models of statistical power-laws in science and engineering \cite{New}-\cite{Arn}. The survival function of a type II Pareto input is $\bar{F}(t)=[1+(t/s)]^{-\alpha}$, where $s$ is a positive scale parameter, and where $\alpha$ is a positive Pareto power. In turn, the density function of a type II Pareto input is $f(t)= \frac{\alpha}{s}[1+(t/s)]^{-\alpha -1}$. 

Substituting the type II Pareto survival and density functions into Eq. (\ref{301}), straightforward calculations imply that: the input's coincidence likelihood is $\gamma = \frac{s}{\alpha \epsilon + \epsilon -1 }(\frac{\alpha}{s})^{\epsilon}$; and the output's coincidence likelihood is $C(\tau)=\gamma \frac{1-[1+(\tau/s)]^{-(\epsilon \alpha + \epsilon -1)}}{1-[1+(\tau/s)]^{-\epsilon \alpha}}$. Consequently, the fact that $\epsilon>1$ further implies that the output's coincidence likelihood is larger than the input's coincidence likelihood, $C\left(\tau \right) >\gamma $. So, for inputs that are Pareto type II, the following conclusion is attained: sharp restart, with any timer $\tau $, decreases diversity.

\begin{figure}[t!]
\centering
\includegraphics[width=8cm]{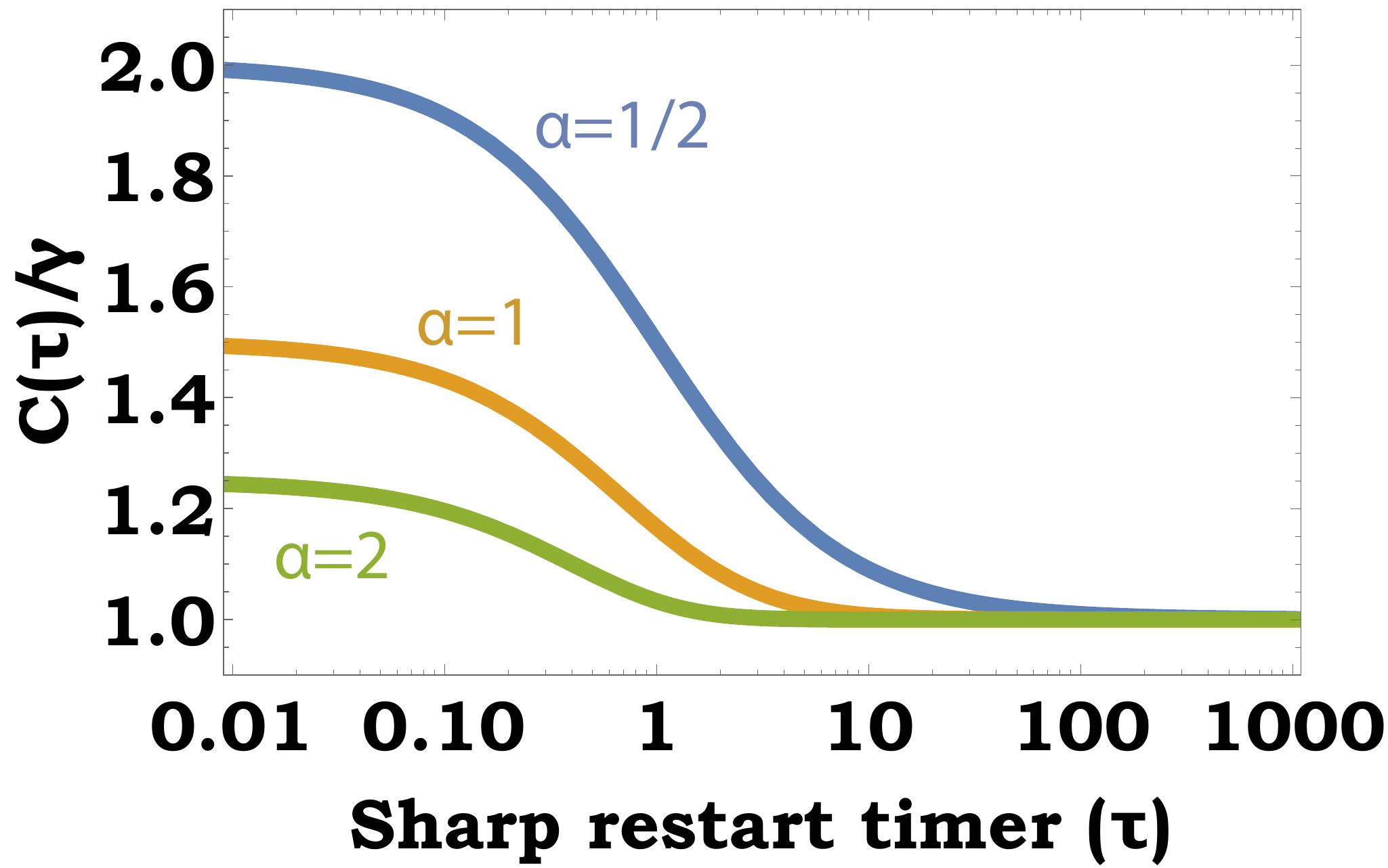}
\caption{An example of a type II Pareto input. In this example the input's survival function is $\bar{F}(t)=[1+(t/s)]^{-\alpha}$, where $s$ is a positive scale parameter, and where $\alpha$ is a positive Pareto power. For this example, the  coincidence-likelihood ratio $C(\tau)/\gamma$ is plotted vs. the sharp-restart timer $\tau$. The plots are for the Pareto powers $\alpha=0.5, 1, 2$; in all three plots the scale parameter is $s=1$, and the Renyi exponent is $\epsilon=2$. Evidently, in all three plots the output's coincidence likelihood $C(\tau)$ is greater than the input's coincidence likelihood $\gamma$, and hence: sharp restart, with any timer $\tau $, decreases diversity.} \label{Pareto}
\end{figure}

The example of a type II Pareto input is depicted in Figure \ref{Pareto}. For this example, the above calculations yield a certain reciprocal structure of the output's coincidence likelihood. As shall now be shown, this reciprocal structure is not unique to the type II Pareto example. Rather, the structure holds for any input, and it gives rise to general `survival criteria' that determine if sharp restart decreases or increases diversity.

\subsection{Survival criteria}

\noindent Eq. (\ref{301}) conceals two statistical distributions that are induced by the input's statistical distribution, and that are defined over the positive half-line. We shall now describe these statistical distributions, and thereafter re-formulate Eq. (\ref{301}) in their terms. This re-formulation will yield general `survival criteria' that shall determine -- for any given input $T$, and for any given timer $\tau$ -- if sharp restart decreases or increases diversity.

The first statistical distribution is governed by the following density
function:
\begin{equation}
\psi \left( t\right) =\frac{1}{\gamma }f\left( t\right) ^{\epsilon }
\label{303}
\end{equation}
($t>0$). Evidently, as the input's density function $f\left( t\right) $ is positive-valued over the positive half-line, so is the\ function $\psi \left( t\right) $. Moreover, the very definition of the coincidence likelihood $\gamma$ implies that the function $\psi \left( t\right) $ has a unit mass over the positive half-line: $\int_{0}^{\infty }\psi \left( t\right) dt=1$. Hence, over the positive half-line, $\psi \left( t\right) $ is indeed a density function. We denote by $\Psi \left( t\right) =\int_{0}^{t}\psi \left( s\right) ds$ ($t\geq 0$) the corresponding distribution function, and by $\bar{\Psi} \left( t\right) =\int_{t}^{\infty }\psi \left( s\right) ds$ ($t\geq 0$) the corresponding survival function.

For an integer Renyi exponent ($\epsilon =2,3,\cdots $) the density function $\psi \left( t\right) $ has the following probabilistic meaning. Consider a sample $\left\{ T_{1},\cdots ,T_{\epsilon }\right\} $ comprising $\epsilon $ IID copies of the random variable $T$. Then -- provided that the coincidence event $T_{1}=\cdots =T_{\epsilon }$ occurred -- $\psi \left( t\right) $ is the likelihood that the common value attained by the IID copies is $t$.

The second statistical distribution is governed by the following survival function:
\begin{equation}
\bar{\Phi}\left( t\right) =\bar{F}\left( t\right) ^{\epsilon }  \label{304}
\end{equation}
($t\geq 0$). Evidently, as the input's survival function $\bar{F}\left(t\right) $ is monotone decreasing from the unit-level $\lim_{t\rightarrow 0} \bar{F}\left( t\right) =1$ to the zero-level $\lim_{t\rightarrow \infty } \bar{F}\left( t\right) =0$, so does the function $\bar{\Phi}\left( t\right) $. Hence, over the positive half-line, $\bar{\Phi}\left( t\right) $ is indeed a survival function. We denote by $\Phi \left( t\right) =1-\bar{F} \left( t\right) ^{\epsilon }$ ($t\geq 0$) the corresponding distribution function, and by $\phi \left( t\right) =\epsilon \bar{F}\left( t\right)^{\epsilon -1}f\left( t\right) $ ($t>0$) the corresponding density function.

For an integer Renyi exponent ($\epsilon =2,3,\cdots $) the survival function $\bar{\Phi}\left( t\right) $ has the following probabilistic meaning. As above, consider a sample $\left\{ T_{1},\cdots ,T_{\epsilon }\right\}$ comprising $\epsilon $ IID copies of the random variable $T$. Then, $\bar{\Phi}\left( t\right) $ is the survival function of the sample's minimum, $\min \left\{ T_{1},\cdots ,T_{\epsilon }\right\} $. Namely: $\bar{\Phi} \left( t\right) =\Pr \left( \min \left\{ T_{1},\cdots ,T_{\epsilon }\right\} >t\right) $.

With the two statistical distributions that were introduced, we can now re-formulate Eq. (\ref{301}). To that end we divide both sides of Eq. (\ref{301}) by the input's coincidence likelihood $\gamma $, and then use the above distribution and survival functions. This yields the following formula for the ratio of the output's coincidence likelihood to the input's coincidence likelihood:

\begin{equation}
\frac{C\left( \tau \right) }{\gamma }=\frac{\Psi \left( \tau \right) }{\Phi\left( \tau \right) }=\frac{1-\bar{\Psi}\left( \tau \right) }{1-\bar{\Phi}\left( \tau \right) }.  \label{305}
\end{equation}

Recalling the inverse relation between the diversity and the coincidence likelihood, Eq. (\ref{305}) straightforwardly implies the following pair of \emph{survival criteria}: 

\begin{enumerate}
\item[$\bullet $] Sharp restart with timer $\tau $ decreases diversity if and only if $\bar{\Psi}\left( \tau \right) <\bar{\Phi}\left(\tau \right) $.

\item[$\bullet $] Sharp restart with timer $\tau $ increases diversity if and only if $\bar{\Psi}\left( \tau \right) >\bar{\Phi}\left(\tau \right) $.
\end{enumerate}

The survival criteria determine if sharp restart decreases/increases diversity by comparing the survival functions of the two statistical distributions that were introduced above. We emphasize that, in general, the survival criteria may provide different answers for different timers $\tau $. Namely, it may be that while sharp restart decreases diversity for some timers $\tau $, it increases diversity for other timers $\tau $ (and vice versa).

As an illustrative demonstration of the survival criteria, we re-visit the above example of a Pareto type II input. Recall that in this example the input's survival and density functions are, respectively, $\bar{F}(t)=[1+(t/s)]^{-\alpha}$ and $f(t)= \frac{\alpha}{s}[1+(t/s)]^{-\alpha -1}$ (where $s$ is a positive scale parameter, and where $\alpha$ is a positive Pareto power). Consequently, straightforward calculations imply that $\bar{\Psi}\left( \tau \right)=[1+(\tau/s)]^{-(\epsilon \alpha + \epsilon -1)}$ and $\bar{\Phi}\left(\tau \right)=[1+(\tau/s)]^{-\epsilon \alpha}$. Note that these survival functions are type II Pareto with scale parameter $s$, and with Pareto powers $\epsilon \alpha + \epsilon -1$ and $\epsilon \alpha$ (respectively). The fact that $\epsilon>1$ implies that $\bar{\Psi}\left( \tau \right) <\bar{\Phi}\left(\tau \right)$ (see Figure \ref{Pareto}). So, the aforementioned conclusion regarding inputs that are Pareto type II is re-attained: sharp restart, with any timer $\tau $, decreases diversity.

\subsection{Diversity invariance}

To conclude this section, we address the following question: are there inputs that sharp restart neither increases nor decreases their diversity? Specifically, inputs that yield the `flat' output diversity $D(\tau)=\delta$ for all timers $\tau $ -- which is tantamount to the `flat' output coincidence-likelihood $C\left( \tau \right) =\gamma $ for all timers $\tau $. 

On the one hand, Eq. (\ref{305}) implies that $C\left( \tau \right) =\gamma $ for all timers $\tau $ if and only if the two statistical distributions -- governed, respectively, by the density functions $\psi \left( t\right) $ and $\phi \left( t\right) $ -- are identical. On the other hand, it is straightforward to check that the density functions $\psi \left( t\right) $ and $\phi \left( t\right) $ are identical if and only if the input's density function is that of the \emph{Exponential distribution}: $f\left( t\right) = \frac{1}{s} \exp(-t/s) $ ($t>0$), where $s $ is a positive scale parameter. Thus, combining these two facts together, we arrive at the following diversity conclusions:

\begin{enumerate}

\item[$\bullet $] The output's diversity is identical to the input's diversity, $D\left( \tau \right) =\delta $ for all timers $\tau $, if and only if the input $T$ is Exponentially-distributed.

\item[$\bullet $] If the input $T$ is not Exponentially-distributed then there exist timers $\tau $ with which the output's diversity differs from the input's diversity, $D\left( \tau \right) \neq \delta $.

\end{enumerate}

\noindent  Note that the diversity of an Exponentially-distributed input is $s \cdot \delta_{\exp}$, where $s$ is the aforementioned scale parameter, and where $\delta_{\exp}=\epsilon ^{1/(\epsilon -1)}$ is the diversity of the unit-scale Exponential distribution.\footnote{Namely, the unit-scale Exponential distribution is characterized by the exponential density function $f(t)=\exp(-t)$ ($t>0$).} The diversity value $\delta_{\exp}$ will play a key role in sections \ref{5} and \ref{6} below.

\section{\label{4}Existence results}

The pair of survival criteria established in the previous section are timer-specific. Namely, for a given timer $\tau $, the pair of survival criteria determine if sharp restart with that specific timer decreases or increases diversity. In this section we shift from timer-specific criteria to \emph{existence criteria}: results that determine the very existence of timers with which sharp restart decreases or increases diversity. As the results of the previous section, the results of this section shall also be based on the statistical distributions that were defined via Eqs. (\ref{303}) and (\ref{304}).

\subsection{Existence criteria I}

The mean of a statistical distribution that is defined over the positive half-line is given by the integral of its survival function. Hence, the mean of the statistical distribution that is governed by the density function $\psi \left( t\right) $ is $\mu _{\psi }=\int_{0}^{\infty }\bar{\Psi}\left(t\right) dt$. Also, the mean of the statistical distribution that is governed by the density function $\phi \left( t\right) $ is $\mu _{\phi}=\int_{0}^{\infty }\bar{\Phi}\left( t\right) dt$. Consequently, we have
\begin{equation}
\int_{0}^{\infty }\left[ \bar{\Psi}\left( t\right) -\bar{\Phi}\left(
t\right) \right] dt=\mu _{\psi }-\mu _{\phi }.  \label{401}
\end{equation}

Note that if the integral appearing in Eq. (\ref{401}) is negative then its integrand -- the difference $\left[ \bar{\Psi}\left( \tau\right) -\bar{\Phi}\left(\tau\right) \right]$ -- must be negative for some $\tau$. Similarly, if the integral is positive then its integrand must be positive for some $\tau$. Thus, combining Eq. (\ref{401}) together with the survival criteria of section \ref{3} yields the following pair of \emph{existence criteria}:

\begin{enumerate}
\item[$\bullet $] If $\mu _{\psi }<\mu _{\phi }$ then there exist timers $\tau $ with which sharp restart decreases diversity.

\item[$\bullet $] If $\mu _{\psi }>\mu _{\phi }$ then there exist timers $\tau $ with which sharp restart increases diversity.
\end{enumerate}

The existence criteria determine if sharp restart decreases/increases diversity by comparing the means of the two statistical distributions that were introduced in the previous section: the one governed by the density function of Eq. (\ref{303}), and the one governed by the survival function of Eq. (\ref{304}). While the survival criteria of the previous section required full knowledge of the two statistical distributions, the existence criteria of this section require knowing only the means of the two statistical distributions.

We note that an alternative way of obtaining these existence criteria is via the following formula:
\begin{equation}
\int_{0}^{\infty }\left[ \frac{C\left( \tau \right) -\gamma }{\gamma}\right] \Phi \left( \tau \right) d\tau =\mu _{\phi }-\mu _{\psi }.
\label{402}
\end{equation}
Eq. (\ref{402}) stems from Eq. (\ref{305}) via the following steps: subtract $1$ from both sides of Eq. (\ref{305}), multiply by the quantity $\Phi \left( \tau \right)$, integrate over positive half-line, and then use Eq. (\ref{401}). 

As an illustrative demonstration of these existence criteria, we revisit the type II Pareto example of the previous section. Recall that in this example the survival function is $\bar{F}(t)=[1+(t/s)]^{-\alpha}$, where $s$ is a positive scale parameter, and where $\alpha$ is a positive Pareto power. Consequently, as noted  above, $\bar{\Psi}\left( \tau \right)=[1+(\tau/s)]^{- (\epsilon \alpha + \epsilon -1)}$ and $\bar{\Phi}\left(\tau \right)=[1+(\tau/s)]^{- \epsilon \alpha}$. Considering the Pareto power $\alpha$ to be greater than one, the corresponding means are $\mu _{\psi }=s/(\epsilon \alpha + \epsilon -2)$ and $\mu _{\phi }=s/(\epsilon \alpha-1)$. Evidently, the fact that $\epsilon>1$ implies that $\mu _{\psi }<\mu _{\phi }$, and hence: there exist timers $\tau $ with which sharp restart decreases diversity.

Regarding the type II Pareto example, the existence conclusion of this section is in full accord with the survival conclusion of the previous section (which asserted that sharp restart, with any timer $\tau$, decreases diversity). As pointed out above: in order to deduce the existence conclusion one needs to know only the means $\mu _{\psi }$ and $\mu _{\phi }$, whereas in order to deduce the survival conclusion one needs to have full information of the survival functions $\bar{\Psi}\left( t\right)$ and $\bar{\Phi}\left(t\right)$.

\subsection{Existence criteria II}

Set $T_{\psi }$ and $T_{\phi }$ to be positive-valued random variables whose statistical distributions are governed, respectively, by the density functions $\psi \left( t\right) $ and $\phi \left( t\right) $. The existence criteria of the previous subsection contested the means of the random variables $T_{\psi }$ and $T_{\phi }$ against each other. In this subsection we shall present existence criteria that contest the random variables $T_{\psi }$ and $T_{\phi }$ directly against each other.

To that end we introduce the density function $\phi _{\max }\left( t\right) =2\Phi \left(t\right) \phi \left( t\right) $ ($t>0$). Specifically, $\phi _{\max }\left(t\right) $ is the density function of the maximum of two IID copies of the random variable $T_{\phi }$. The following formula emanates from Eq. (\ref {305}): 
\begin{equation}
\int_{0}^{\infty }\left[ \frac{C\left( \tau \right) -\gamma }{\gamma }\right] \phi _{\max }\left( \tau \right) d\tau =2\Pr \left( T_{\psi }\leq T_{\phi}\right) -1,  \label{403}
\end{equation}
where the random variables that appear on the right-hand side of Eq. (\ref {403}), $T_{\psi }$ and $T_{\phi }$, are independent of each other. The derivation of Eq. (\ref{403}) is detailed in the Methods. 

Using argumentation which is similar to that following Eq. (\ref{401}), Eq. (\ref {403}) yields the following pair of \emph{existence criteria}: 

\begin{enumerate}
\item[$\bullet $] If $\Pr \left( T_{\psi }\leq T_{\phi }\right) >\frac{1}{2}$ then there exist timers $\tau $ with which sharp restart decreases diversity.

\item[$\bullet $] If $\Pr \left( T_{\psi }\leq T_{\phi }\right) <\frac{1}{2}$ then there exist timers $\tau $ with which sharp restart increases diversity.
\end{enumerate}

\noindent The existence criteria determine if sharp restart decreases/increases diversity by contesting the random variable $T_{\psi }$ against the random variable $T_{\phi }$, and then comparing the probability of the event $T_{\psi }\leq T_{\phi }$ to the `benchmark' probability $\frac{1}{2}$.

Equation (\ref{403}) can be re-formulated in terms of the minimum $\min \left\{ T_{1},T_{2}\right\} $ of two IID copies of the input $T$. Specifically, denoting by $\delta _{\min }=\mathcal{D}_{\epsilon } \left[ \min \left\{ T_{1},T_{2}\right\} \right] $ the minimum's diversity, the re-formulation of Eq. (\ref{403}) is:
\begin{equation}
\int_{0}^{\infty }\left[ \frac{C\left( \tau \right) -\gamma }{\gamma }\right] \phi _{\max }\left( \tau \right) d\tau =\left( \frac{\delta }{2\delta _{\min}}\right) ^{\epsilon -1}-1.  \label{404}
\end{equation}
The derivation of Eq. (\ref{404}) is detailed in the Methods. 

Eq. (\ref{404}) yields the following alternative formulation of this subsection's pair of existence criteria:

\begin{enumerate}
\item[$\bullet $] If $\delta > 2\delta _{\min }$ then there exist timers $\tau $ with which sharp restart decreases diversity.

\item[$\bullet $] If $\delta < 2\delta _{\min }$ then there exist timers $\tau $ with which sharp restart increases diversity.
\end{enumerate}

\noindent In this formulation, the existence criteria determine if sharp restart decreases/increases diversity by comparing two diversities: that of the input $T$ to that of the minimum $\min \left\{ T_{1},T_{2}\right\} $ of two IID copies of the input $T$.

As an illustrative demonstration of these existence criteria, consider an input whose distribution is Weibull. The Weibull distribution \cite{Wei} is one of the three universal laws emanating from the Fisher-Tippett-Gnedenko theorem \cite{FT}-\cite{Gne} of Extreme Value Theory \cite{Gal}-\cite{RT}, and it has numerous uses in science and engineering \cite{MXJ}-\cite{McC}. The survival function of a Weibull input is $\bar{F}\left( t\right) =\exp [-(t/s)^{\alpha }]$, where $s$ is a positive scale parameter, and where $\alpha$ is a positive Weibull exponent. 

The Weibull input displays the following scaling property: $\min \left\{T_{1},T_{2}\right\} =2^{-1/\alpha }T$, where the equality is in law. In turn, this scaling property implies that $\delta _{\min }=2^{-1/\alpha }\delta $. So, when the Weibull exponent is in the `sub-Exponential' range $\alpha <1$ then: $\delta >2\delta _{\min }$, and hence there exist timers $\tau $ with which sharp restart decreases diversity. And, when the Weibull exponent is in the `super-Exponential' range $\alpha >1$ then: $\delta <2\delta _{\min }$, and hence there exist timers $\tau $ with which sharp restart increases diversity.

In the `sub-Exponential' range $\alpha <1$ the Weibull distribution manifests the Stretched Exponential distribution -- which plays a key role in statistical physics \cite{WW}-\cite{CK}. So, in particular, the aforementioned Weibull example yields the following Stretched-Exponential conclusion: there exist timers $\tau $ with which sharp restart can always decrease the diversity of Stretched-Exponential tasks. We shall re-visit the Weibull example, and bolster the above Weibull conclusions, in section \ref{6} below.

\section{\label{5}Asymptotic analysis}

Section \ref{3} analysed sharp restart with general timers $0<\tau<\infty$. In this section shall address two extreme cases of sharp restart. \textbf{A}) The case of `high-frequency resetting' which employs very small timers $\tau\ll1$ -- henceforth termed \emph{fast timers}. \textbf{B}) The case of `low-frequency resetting' which employs very large timers $\tau\gg1$ -- henceforth termed \emph{slow timers}. As noted in the introduction, the case of low-frequency resetting attracted substantial scientific interest \cite{Slow_Time_1}-\cite{Slow_Time_11}.

\subsection{Fast and slow criteria}

The cases of sharp restart with fast and slow timers correspond, respectively, to the timer limits $\tau \rightarrow 0$ and $\tau \rightarrow \infty $. In order to analyze the asymptotic behavior of sharp restart in these limits we shall use the notion of \emph{hazard function} (described below). Also, we shall use the shorthand notation $\varphi \left( 0\right) =\lim_{t\rightarrow 0}\varphi \left( t\right) $ and $\varphi \left( \infty \right)=\lim_{t\rightarrow \infty }\varphi \left( t\right) $ for the limit values of a general positive-valued function $\varphi \left( t\right) $ that is defined over the positive half-line ($t>0$); these limit values are assumed to exist in the wide sense, $0\leq \varphi \left( 0\right) ,\varphi \left( \infty \right) \leq \infty $. 

The input's \emph{hazard function} played a focal role in the entropy analysis \cite{EntSR}, in the mean-performance analysis \cite{MP1SR}, and in the tail-behavior analysis \cite{TBSR} of sharp restart. The input's hazard function is the negative logarithmic derivative of the input's survival function:
\begin{equation}
H\left( t\right) =-\left\{ \ln \left[ \bar{F}\left( t\right) \right]
\right\} ^{\prime }=\frac{f\left( t\right) }{\bar{F}\left( t\right) }.  \label{501}
\end{equation}

\noindent The hazard function of Eq. (\ref{501}) has the following probabilistic meaning: $H\left(t\right) $ is the likelihood that the input be realized right at time $t$, given the information that the input was not realized up to time $t$. The hazard function -- a.k.a. \textquotedblleft hazard rate\textquotedblright\ and \textquotedblleft failure rate\textquotedblright\ -- is a widely applied tool in survival analysis \cite{KP}-\cite{Col} and in reliability engineering \cite{BP}-\cite
{Dhi}.

Now, with the input's hazard function $H\left( t\right) $ at hand, we are all set to analyze the asymptotic behavior of sharp restart. Eqs. (\ref{503})-(\ref{504}) below will incorporate the number
\begin{equation}
\delta_{\exp} =\epsilon ^{1/(\epsilon -1)}.  \label{502}
\end{equation}
As noted at the end of section \ref{3}, the number $\delta_{\exp}$ is the diversity of the unit-scale Exponential distribution. 

Firstly, we address the case of \emph{fast timers} ($\tau\ll1$). Taking the limit $\tau \rightarrow 0$ in the middle part of Eq. (\ref{305}), while using L'Hospital's rule, yields the following limit-value of the output's diversity: 
\begin{equation}
D\left( 0\right) =\frac{\delta_{\exp} }{H\left( 0\right) }. \label{503}
\end{equation}
The derivation of Eq. (\ref{503}) is detailed in the Methods. In turn, Eq. (\ref{503}) implies the following pair of \emph{fast criteria}.

\begin{enumerate}
\item[$\bullet $] If $H\left( 0\right) <\delta_{\exp} /\delta $ then sharp restart with fast timers increases diversity; in particular, this criterion applies whenever the hazard function vanishes at the origin, $H\left(0\right) =0$.

\item[$\bullet $] If $H\left( 0\right) >\delta_{\exp} /\delta $ then sharp restart with fast timers decreases diversity; in particular, this criterion applies whenever the hazard function explodes at the origin, $H\left(0\right) =\infty $.
\end{enumerate}

Secondly, we address the case of \emph{slow timers} ($\tau \gg1$). Taking the limit $\tau \rightarrow \infty $ in the middle part of Eq. (\ref{305}) yields the following limit-value of the output's coincidence likelihood: $C\left( \infty \right) =\gamma $. In turn, a calculation using Eq. (\ref{305}) and L'Hospital's rule asserts that the asymptotic behavior of the output's coincidence likelihood $C\left( \tau \right) $ about its limit value $C\left( \infty \right) =\gamma $ is given by the following limit: 
\begin{equation}
\lim_{\tau \rightarrow \infty }\frac{C\left( \tau \right) -\gamma }{\bar{\Phi}\left( \tau \right) }=\left( \frac{1}{\delta }\right) ^{\epsilon -1}-\left[\frac{1}{\delta_{\exp} }H\left( \infty \right) \right] ^{\epsilon -1}.
\label{504}
\end{equation}
The derivation of Eq. (\ref{504}) is detailed in the Methods. In turn, Eq. (\ref{504}) implies the following pair of \emph{slow criteria}.

\begin{enumerate}
\item[$\bullet $] If $H\left( \infty \right) <\delta_{\exp} /\delta $ then sharp restart with slow timers decreases diversity; in particular, this criterion applies whenever the hazard function vanishes at infinity, $H\left( \infty \right) =0$.

\item[$\bullet $] If $H\left( \infty \right) >\delta_{\exp} /\delta $ then sharp restart with slow timers increases diversity; in particular, this criterion applies whenever the hazard function explodes at infinity, $H\left( \infty \right) =\infty $.
\end{enumerate}

The fast and slow criteria display a common pattern: comparing the limit-values of the hazard function $H(t)$ to the ratio of diversities $\delta_{\exp} /\delta $. This ratio manifests the height of a flat hazard function that characterizes an Exponential distribution whose diversity is equal to the input's diversity $\delta$. Hence, in effect, the ratio of diversities $\delta_{\exp} /\delta $ serves -- in the fast and slow criteria -- as an `Exponential hazard-function benchmark' to which the limit-values of the input's hazard function are compared.

\subsection{Fast and slow timers}

In the previous subsection we addressed the case of \emph{fast timers} ($\tau\ll1$), and the case of \emph{slow timers} ($\tau \gg1$). However, we did not provide the precise meanings of these timers. Namely: how small should the timer $\tau $ be in order to qualify as `fast'? and how large should the timer $\tau $ be in order to qualify as `slow'? Considering the input's hazard function $H\left( t\right) $ to be continuous over the positive half-line ($t>0$), in this subsection we answer these two questions.

The interplay between the following terms assumed a key role in the previous subsection: the input's hazard function $H\left( t\right) $ on the one hand, and the ratio of diversities $\delta_{\exp} /\delta $ on the other hand. This interplay will assume a key role also in this subsection, via the two following thresholds:
\begin{equation}
\tau _{\ast }=\inf \left\{ t>0 \vert H\left( t\right) =\frac{\delta_{\exp} }{\delta }\right\}, \label{601}
\end{equation}
and
\begin{equation}
\tau ^{\ast }=\sup \left\{ t>0 \vert H\left( t\right) =\frac{\delta_{\exp} }{\delta }\right\}. \label{602}
\end{equation}
Namely, the thresholds $\tau _{\ast }$ and $\tau ^{\ast }$ are, respectively, the smallest and largest times at which the hazard function $H\left( t\right)$ intersects the level $\delta_{\exp} /\delta $. In particular, if the hazard function $H\left( t\right)$ does \emph{not} intersect the level $\delta_{\exp} /\delta $ then: $\tau _{\ast }=\infty $ and $\tau ^{\ast }=0$.

A calculation based on Eq. (\ref{305}) yields the following formulae: 
\begin{equation}
C\left( \tau \right) -\gamma =\frac{1}{\epsilon \Phi \left( \tau \right) }\int_{0}^{\tau }\left[ H\left( t\right) ^{\epsilon -1}-\left( \frac{\delta_{\exp} }{\delta }\right) ^{\epsilon -1}\right] \phi \left( t\right) dt .
\label{603}
\end{equation}
and
\begin{equation}
C\left( \tau \right) -\gamma =\frac{1}{\epsilon \Phi \left( \tau \right) }\int_{\tau }^{\infty }\left[ \left( \frac{\delta_{\exp} }{\delta }\right)^{\epsilon -1}-H\left( t\right) ^{\epsilon -1}\right] \phi \left( t\right) dt.  \label{604}
\end{equation}
The derivations of Eqs. (\ref{603}) and (\ref{604}) are detailed in the Methods. In Eqs. (\ref{603}) and (\ref{604}), the input's hazard function $H(t)$ is compared to the level $\delta_{\exp} /\delta $. As noted at the close of the previous subsection, the level $\delta_{\exp} /\delta $ manifests an `Exponential hazard-function benchmark'.

Consider the case of \emph{fast timers}: $\tau \ll1$. The fast-timer criteria of the previous subsection, combined together with Eq. (\ref{603}), straightforwardly yield the two following conclusions. \textbf{I}) If $H\left(0\right) <\delta_{\exp} /\delta $ then sharp restart increases diversity for all timers $\tau $ in the range $\tau <\tau _{\ast }$. \textbf{II}) If $H\left( 0\right)>\delta_{\exp} /\delta $ then sharp restart decreases diversity for all timers $\tau $ in the range $\tau <\tau _{\ast }$. Hence, the threshold $\tau _{\ast}$ of Eq. (\ref{601}) defines the \emph{range} of \emph{fast timers}.

Consider the case of \emph{slow timers}: $\tau \gg1$. The slow-timer criteria of the previous subsection, combined together with Eq. (\ref{604}), straightforwardly yield the two following conclusions. \textbf{I}) If $H\left(\infty \right) <\delta_{\exp} /\delta $ then sharp restart decreases diversity for all timers $\tau $ in the range $\tau >\tau ^{\ast }$. \textbf{II}) If $H\left(\infty \right) >\delta_{\exp} /\delta $ then sharp restart increases diversity for all timers $\tau $ in the range $\tau >\tau ^{\ast }$. Hence, the threshold $\tau ^{\ast }$ of Eq. (\ref{602}) defines the \emph{range} of \emph{slow timers}.

\begin{figure}[t!]
\centering
\includegraphics[width=8cm]{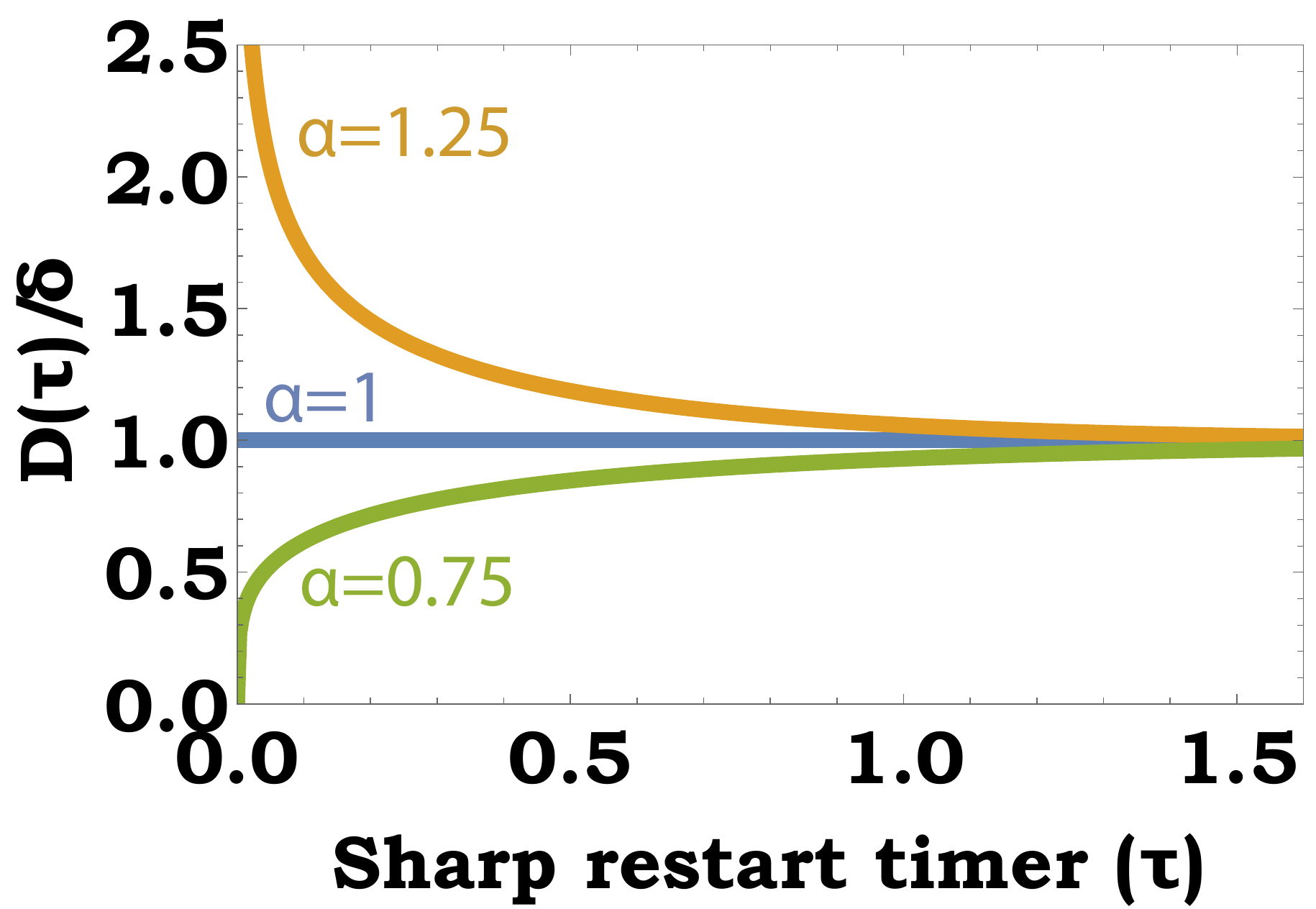}
\caption{An example of a Weibull input. In this example the input's survival function is $\bar{F}\left( t\right) =\exp [-(t/s)^{\alpha }]$, where $s$ is a positive scale parameter, and where $\alpha$ is a positive Weibull exponent. For this example, the ratio of diversities $D(\tau)/ \delta$ is plotted vs. the sharp-restart timer $\tau$. The plots are for the Weibull exponents $\alpha=0.75, 1, 1.25$; in all three plots the scale parameter is $s=1$, and the Renyi exponent is $\epsilon=2$. On the one hand, in the case of the `sub-Exponential' Weibull exponent $\alpha=0.75$: sharp restart, with any timer $\tau $, decreases diversity. On the other hand, in the case of the `super-Exponential' Weibull exponent $\alpha=1.25$: sharp restart, with any timer $\tau $, increases diversity. In the case of the `Exponential' Weibull exponent $\alpha=1$: sharp restart neither decreases nor increases diversity.} \label{WeibullFig}
\end{figure}

As an illustrative demonstration of the ranges of fast and slow timers, let us re-visit the Weibull example of section \ref{4}. Recall that in this example the survival function is $\bar{F}\left( t\right) =\exp [-(t/s)^{\alpha }]$, where $s$ is a positive scale parameter, and where $\alpha$ is a positive Weibull exponent. Consequently, the hazard function is $H(t)=ct^{\alpha-1}$, with coefficient $c=\alpha s^{-\alpha}$.

When the Weibull exponent is in the `sub-Exponential' range $\alpha <1$ then the hazard function is monotone decreasing from $H(0)=\infty$ to $H(\infty)=0$. Hence, the results of the previous subsection assert that sharp restart with either fast or slow timers decreases diversity. The shape of the hazard function implies that it intersects the level $\delta_{\exp} /\delta $ at the unique threshold $\tau _{\ast}=\tau ^{\ast}=[c \delta / \delta_{\exp}]^{1/(1-\alpha)}$. Consequently, Eqs. (\ref{603}) and (\ref{604}) assert that: sharp restart with \emph{any} timer decreases diversity (see Figure \ref{WeibullFig}).

When the Weibull exponent is in the `super-Exponential' range $\alpha >1$ the hazard function is monotone increasing from $H(0)=0$ to $H(\infty)=\infty$. Hence, the results of the previous subsection assert that sharp restart with either fast or slow timers increases diversity. The shape of the hazard function implies that it intersects the level $\delta_{\exp} /\delta $ at the aforementioned unique threshold ($\tau _{\ast}=\tau ^{\ast}=[c \delta / \delta_{\exp}]^{1/(1-\alpha)}$). Consequently, Eqs. (\ref{603}) and (\ref{604}) assert that: sharp restart with \emph{any} timer increases diversity (see Figure \ref{WeibullFig}).

In fact, the above conclusion regarding the Weibull example in the `sub-Exponential' range holds for any input whose hazard function is continuous and monotone decreasing from $H(0)=\infty$ to $H(\infty)=0$. And, the above conclusion regarding the Weibull example in the `super-Exponential' range holds for any input whose hazard function is continuous and monotone increasing from $H(0)=0$ to $H(\infty)=\infty$.

\section{\label{6}Joint mean-diversity perspective}

The fast and slow criteria developed in the previous section are `diversity analogues' of fast and slow criteria for the mean-performance of sharp restart \cite{MP1SR}. We now turn to intertwine the fast and slow criteria established here together with the fast and slow criteria established in \cite{MP1SR} -- doing so in order to obtain a joint mean-diversity perspective of sharp restart.

Applying sharp restart, the `input' task-completion time $T$ is mapped to the `output' task-completion time $T_{R}$. The studies \cite{MP1SR}-\cite{MP2SR} examined the effect of sharp restart on mean-performance: is the output's mean smaller or larger than the input's mean? In other words, does the mean task-completion time decrease $\mathbf{E}\left[ T_R\right]<\mathbf{E}\left[ T\right]$ in response to sharp restart, or does it increase $\mathbf{E}\left[ T_R\right]>\mathbf{E}\left[ T\right]$?

In what follows we use the shorthand $\mu=\mathbf{E}\left[ T\right]$ to denote the input's mean. For fast and slow timers, the following pairs of criteria were established with regard to the effect of sharp restart on mean-performance \cite{MP1SR}. \emph{Fast criteria}: if $H\left( 0\right) < 1/\mu $ then the mean increases; and if $H\left( 0\right) > 1/\mu $ then the mean decreases. \emph{Slow criteria}: if $H\left( \infty \right) < 1/\mu $ then the mean decreases; and if $H\left(  \infty \right) > 1/\mu $ then the mean increases.

Note that the reciprocal of the input's mean $1/\mu$ can be re-written as the ratio $\mu_{\exp} / \mu$, where: $\mu_{\exp}=1$ is the mean of the unit-scale Exponential distribution. Hence, the fast and slow criteria regarding mean-performance display the following common pattern: comparing the limit values of the hazard function $H(t)$ to the ratio of means $\mu_{\exp} / \mu$. This pattern is very similar to the aforementioned pattern that the fast and slow criteria regarding diversity display: comparing the limit values of the hazard function $H(t)$ to the  ratio of diversities $\delta_{\exp} /\delta $.

\begin{figure}[t!]
\centering
\includegraphics[width=10cm]{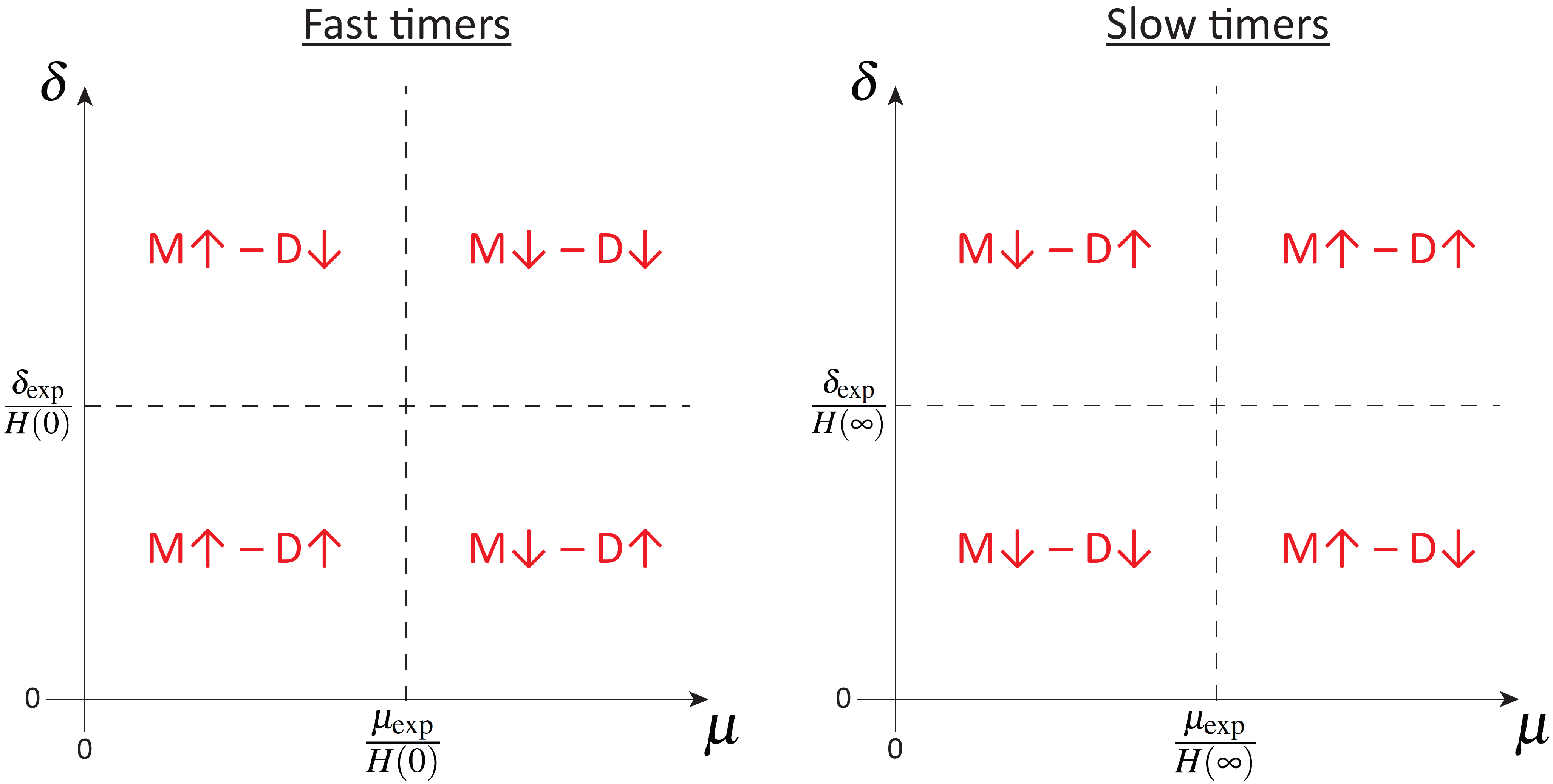}
\caption{Phase-diagram representations of the mean-diversity classifications for sharp restart with fast timers (Table 1) and with slow timers (Table 2). The phase space comprises a horizontal axis that manifests the input's mean $\mu $, and a vertical axis that manifests the input's diversity $\delta $. Rectilinear boarders separate the phase space to four regions that correspond to the four mean-diversity scenarios: mean decrease and diversity decrease ($M\downarrow$-$D\downarrow$); mean decrease and diversity increase ($M\downarrow$-$D\uparrow$); mean increase and diversity decrease ($M\uparrow$-$D\downarrow$); and mean increase and diversity increase ($M\uparrow$-$D\uparrow$). The rectilinear boarders are based on the numbers $\mu_{\exp}=1$ and $\delta_{\exp}=\epsilon^{1/(\epsilon -1)}$ (the mean and the diversity of the unit-scale Exponential distribution), and on the limit-values of the input's hazard function: the limit-value $H\left( 0\right) $ in the case of fast timers; and the limit-value $H\left( \infty \right) $ in the case of slow timers.}
\label{phase-diagram1}
\end{figure}

When applying a given sharp restart algorithm, there are four possible mean-diversity scenarios. On the one hand there are two scenarios in which the algorithm has an aligned effect on mean-performance and on diversity: mean decrease and diversity decrease ($M\downarrow$-$D\downarrow$); mean increase and diversity increase ($M\uparrow$-$D\uparrow$). On the other hand there are two scenarios in which the algorithm has an antithetical effect on mean-performance and on diversity: mean decrease and diversity increase ($M\downarrow$-$D\uparrow$); mean increase and diversity decrease ($M\uparrow$-$D\downarrow$). The similar patterns of the fast and slow criteria yield neat classifications -- presented in Tables 1 and 2 -- of the four mean-diversity scenarios. Alternative representations of these classifications, via mean-diversity phase diagrams, are presented in Figure \ref{phase-diagram1}.

\newpage

\bigskip\

\begin{center}
{\LARGE Table 1}
\end{center}
\
\begin{center}
\begin{tabular}{|l|l|l|}
\hline
& $%
\begin{array}{c}
\\ 
\multicolumn{1}{l}{\textbf{Diversity}} \\ 
\textbf{Decrease} \\ 
\textbf{\ }%
\end{array}%
$ & $%
\begin{array}{l}
\textbf{\ } \\ 
\textbf{Diversity} \\ 
\textbf{Increase} \\ 
\textbf{\ }%
\end{array}%
$ \\ \hline
$%
\begin{array}{c}
\textbf{\ } \\ 
\multicolumn{1}{l}{\textbf{Mean}} \\ 
\textbf{Decrease} \\ 
\textbf{\ }%
\end{array}%
$ & $H\left( 0\right) >\frac{\mu_{\exp}}{\mu },\frac{\delta_{\exp} }{\delta }$ & $\frac{\mu_{\exp}}{\mu }%
<H\left( 0\right) <\frac{\delta_{\exp} }{\delta }$ \\ \hline
$%
\begin{array}{c}
\textbf{\ } \\ 
\multicolumn{1}{l}{\textbf{Mean}} \\ 
\textbf{Increase} \\ 
\textbf{\ }%
\end{array}%
$ & $\frac{\delta_{\exp} }{\delta }<H\left( 0\right) <\frac{\mu_{\exp}}{\mu }$ & $H\left(
0\right) <\frac{\mu_{\exp}}{\mu },\frac{\delta_{\exp} }{\delta }$ \\ \hline
\end{tabular}
\end{center}

\bigskip\ 

\textbf{Table 1}: Classification of the four mean-diversity scenarios in the case of fast timers ($\tau \ll1$). The classification uses the numbers $\mu_{\exp}=1$ and $\delta_{\exp}=\epsilon^{1/(\epsilon -1)}$ (the mean and the diversity of the unit-scale Exponential distribution), and the following features of the input: its mean $\mu $; its diversity $\delta $; and the limit-value $H\left( 0\right) $ of its hazard function. For example, if $H\left( 0\right) >\mu_{\exp} / \mu$ and $H\left( 0\right) >\delta_{\exp} / \delta$, then (when sharp restart with fast timers is applied): both the mean and the diversity of the output will be smaller than that of the input.

\bigskip\ 

\begin{center}
{\LARGE Table 2}
\end{center}
\ 
\begin{center}
\begin{tabular}{|l|l|l|}
\hline
& $%
\begin{array}{c}
\textbf{\ } \\ 
\multicolumn{1}{l}{\textbf{Diversity}} \\ 
\textbf{Decrease} \\ 
\textbf{\ }%
\end{array}%
$ & $%
\begin{array}{l}
\textbf{\ } \\ 
\textbf{Diversity} \\ 
\textbf{Increase} \\ 
\textbf{\ }%
\end{array}%
$ \\ \hline
$%
\begin{array}{c}
\textbf{\ } \\ 
\multicolumn{1}{l}{\textbf{Mean}} \\ 
\textbf{Decrease} \\ 
\textbf{\ }%
\end{array}%
$ & $H\left( \infty \right) <\frac{\mu_{\exp}}{\mu },\frac{\delta_{\exp} }{\delta }$ & $\frac{%
\delta_{\exp} }{\delta }<H\left( \infty \right) <\frac{\mu_{\exp}}{\mu }$ \\ \hline
$%
\begin{array}{c}
\textbf{\ } \\ 
\multicolumn{1}{l}{\textbf{Mean}} \\ 
\textbf{Increase} \\ 
\textbf{\ }%
\end{array}%
$ & $\frac{\mu_{\exp}}{\mu }<H\left( \infty \right) <\frac{\delta_{\exp} }{\delta }$ & $H\left(
\infty \right) >\frac{\mu_{\exp}}{\mu },\frac{\delta_{\exp} }{\delta }$ \\ \hline
\end{tabular}
\end{center}

\bigskip\ 

\textbf{Table 2}: Classification of the four mean-diversity scenarios in the case of slow timers ($\tau \gg1$). The classification uses the numbers $\mu_{\exp}=1$ and $\delta_{\exp}=\epsilon^{1/(\epsilon -1)}$ (the mean and the diversity of the unit-scale Exponential distribution), and the following features of the input: its mean $\mu $; its diversity $\delta $; and the limit-value $H\left( \infty \right) $ of its hazard function. For example, if $H\left( \infty\right) <\mu_{\exp} / \mu$ and $H\left( \infty\right) < \delta_{\exp} / \delta$, then (when sharp restart with slow timers is applied): both the mean and the diversity of the output will be smaller than that of the input.

\newpage

In order to demonstrate the four mean-diversity scenarios, we end this section with a Pareto-Exponential example: an input with power-law statistics up to the time epoch $t=1$, and with exponential statistics afterward. Specifically, the input's density function is $f\left( t\right) =\frac{1}{2}at^{a-1}$ over the temporal interval $0<t\leq 1$, where $a$ is a positive Pareto power. And the input's density function is $f\left( t\right) =\frac{1}{2}b \exp [-b(t-1)] $ over the temporal ray $1<t<\infty $, where $b$ is a positive Exponential `rate'.

In this example the input's mean is $\mu =1-\frac{1}{2a+2}+\frac{1}{2b}$. With regard to the diversity, we set the Renyi exponent $\epsilon =2$ (which corresponds to Simpson's index, to Hirschman's index, and to the participation ratio).  Consequently, in this example the input's diversity is finite if and only if the Pareto power is in the range $a>\frac{1}{2}$ -- in which case $\delta_{\exp} / \delta =2\gamma =\frac{1}{4}(\frac{2a^{2}}{2a-1}+b)$.

The limit-value $H\left( 0\right) $ of the input's hazard function depends on the Pareto power as follows: $H\left( 0\right) =\infty$ when $a<1$ and $H\left( 0\right) =0$ when $a>1$. So, for $a\neq 1$, sharp restart with fast timers yields the following mean-diversity scenarios: mean decrease and diversity decrease ($M\downarrow$-$D\downarrow$) when $a<1$; and mean increase and diversity increase ($M\uparrow$-$D\uparrow$) when $a>1$. When $a=1$ then: the limit-value of the input's hazard function is $H\left( 0\right) =\frac{1}{2}$; the input's mean is $\mu =\frac{3}{4}+\frac{1}{2b}$; and $\delta_{\exp} / \delta  =\frac{1}{2}+\frac{1}{4}b$. Consequently, for $a=1$, sharp restart with fast timers yields the following mean-diversity scenarios: mean decrease and diversity increase ($M\downarrow$-$D\uparrow$) when $b<\frac{2}{5}$; mean increase and diversity increase ($M\uparrow$-$D\uparrow$) when $b>\frac{2}{5}$.

\begin{figure}[t!]
\centering
\includegraphics[width=8cm]{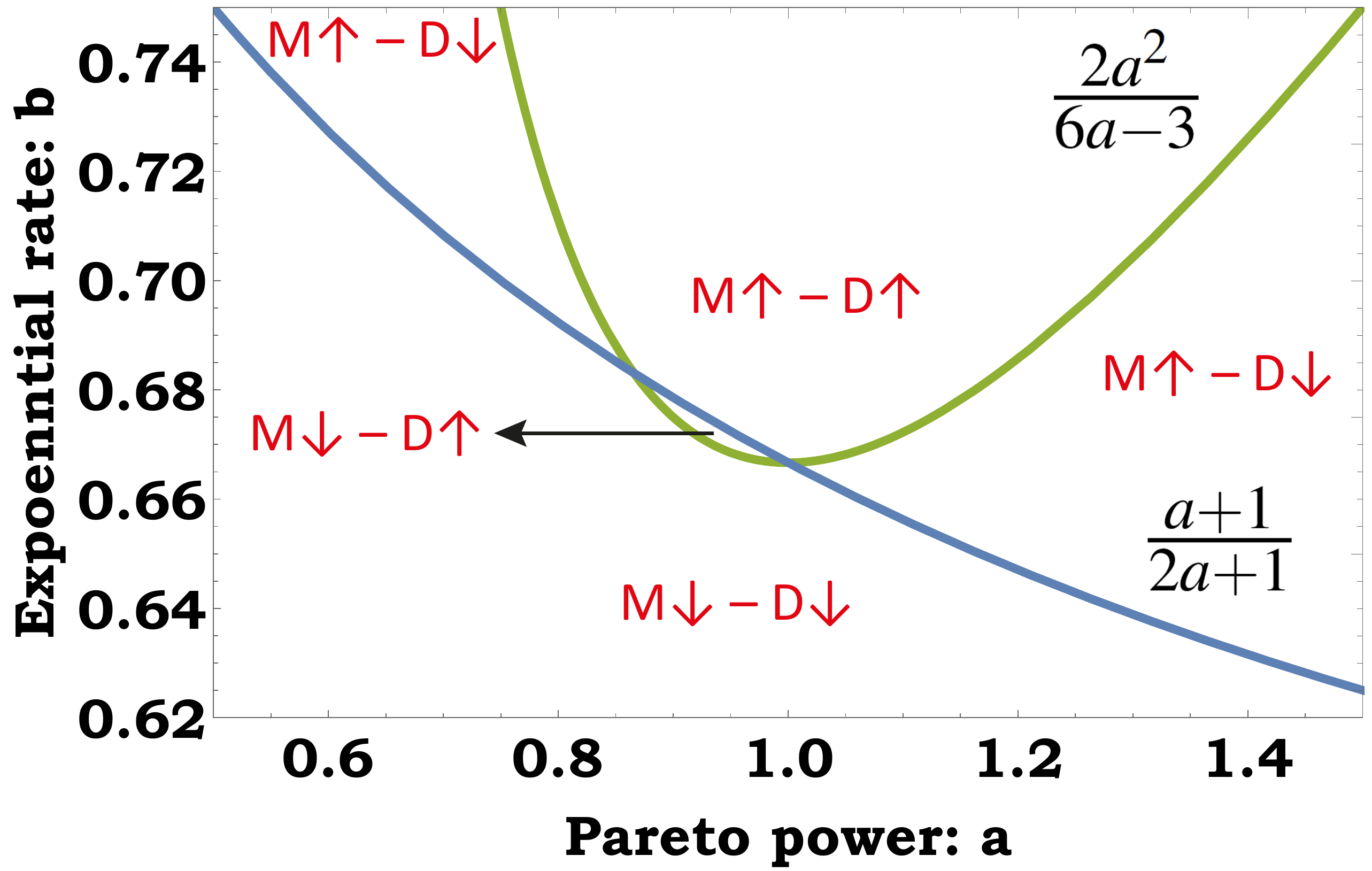}
\caption{Phase-diagram classification -- for the Pareto-Exponential example -- of the four mean-diversity scenarios in the case of slow timers ($\tau \gg1$). The diversity's Renyi exponent is set to be $\epsilon =2$. The horizontal axis manifests the example's Pareto power $a>\frac{1}{2}$, and the vertical axis manifests the example's Exponential rate $b>0$. The phase-diagram involves two curves: the blue monotone decreasing curve $b=\frac{a+1}{2a+1}$ (which emanates from equating the limit-value $H(\infty)$ to the ratio $\mu_{exp}/{\mu}$); and the green U-shaped curve $b=\frac{2a^{2}}{6a-3}$ (which emanates from equating the limit-value $H(\infty)$ to the ratio $\delta_{\exp} / \delta$). The blue and green curves intersect at $a=\frac{1}{2}\sqrt{3}$ and at $a=1$. The four mean-diversity scenarios are as follows. Mean decrease and diversity decrease ($M\downarrow$-$D\downarrow$): the region below both curves. Mean increase and diversity increase ($M\uparrow$-$D\uparrow$): the region above both curves. Mean increase and diversity decrease ($M\uparrow$-$D\downarrow$): the region between the curves, for $a<\frac{1}{2}\sqrt{3}$; and the region between the curves, for $a>1$. Mean decrease and diversity increase ($M\downarrow$-$D\uparrow$): the region between the curves, for $\frac{1}{2}\sqrt{3}<a<1$. } 
\label{phase-diagram3}
\end{figure}

The input's hazard function yields the limit-value $H\left( \infty \right) =b$. In turn, slow timers lead to all four mean-diversity scenarios. These four scenarios are depicted, in Figure \ref{phase-diagram3}, via a phase diagram that corresponds to the two parameters of the Pareto-Exponential example: the Pareto power $a$, and the Exponential rate $b$. Note how the simple rectilinear structure of the general phase-diagram of Figure \ref{phase-diagram1} (which is parameterized by the mean $\mu $ and by the diversity $\delta $) shifts to the intricate non-linear structure of the specific phase-diagram of Figure \ref{phase-diagram3} (which is parameterized by the parameters of the Pareto-Exponential example).

\bigskip\

\section{\label{7}Summary}

This paper presented a comprehensive diversity analysis of the \emph{sharp-restart algorithm}. The algorithm is a non-linear map that receives an input $T$, a positive-valued random variable that manifests the time it takes to accomplish a given task of interest. Using a positive timer parameter $\tau $, the algorithm restarts the task -- as long as it is not accomplished -- every $\tau$ time-units. Doing so, the algorithm produces an output $T_{R}$, a positive-valued random variable that manifests the time it takes to accomplish the task \emph{under sharp restart}.

The analysis focused on comparing the output's diversity $D\left( \tau \right) $ (which is a function of the timer $\tau $) to the input's diversity $\delta $. In turn, this comparison established a detailed diversity roadmap for the sharp-restart algorithm: five pairs of analytic criteria that determine when the application of sharp restart decreases diversity, $D\left( \tau \right) <\delta $; and when it increases diversity, $D\left(\tau \right) >\delta $. These pairs of criteria are summarized in Table 3. The first three pairs of criteria use two statistical distributions that are induced by the input's statistical distribution: one governed by the density function $\psi \left( t\right) $ (see Eq. (\ref{303})), and the other governed by the density function $\phi \left( t\right)$ (see Eq. (\ref{304})). The last two pairs of criteria use the limit-values, $H(0)$ and $H(\infty)$, of the input's hazard function $H(t)$ (see Eq. (\ref{501})).

\newpage 

\begin{center}

{\LARGE Table 3 }

\ \ \ 

\begin{tabular}{|l|l|l|l|l|}
\hline
& $%
\begin{array}{c}
\text{\textbf{\ }} \\ 
\text{\textbf{Timer}} \\ 
\text{\textbf{\ }}%
\end{array}%
$ & $%
\begin{array}{c}
\text{\textbf{\ }} \\ 
\text{\textbf{Parameter}} \\ 
\text{\textbf{\ }}%
\end{array}%
$ & $%
\begin{array}{c}
\text{\textbf{\ }} \\ 
\text{\textbf{Decrease}} \\ 
\text{\textbf{\ }}%
\end{array}%
$ & $%
\begin{array}{c}
\text{\textbf{\ }} \\ 
\text{\textbf{Increase}} \\ 
\text{\textbf{\ }}%
\end{array}%
$ \\ \hline
\textbf{I} & $%
\begin{array}{c}
\text{ } \\ 
\text{General} \\ 
\text{ }%
\end{array}%
$ & $0<\tau <\infty $ & $\bar{\Psi}\left( \tau \right) <\bar{\Phi}\left(
\tau \right) $ & $\bar{\Psi}\left( \tau \right) >\bar{\Phi}\left( \tau
\right) $ \\ \hline
\textbf{II} & $%
\begin{array}{c}
\text{ } \\ 
\text{Existence} \\ 
\text{ }%
\end{array}%
$ & --------- & $\mu _{\psi }<\mu _{\phi }$ & $\mu _{\psi }>\mu _{\phi }$ \\ 
\hline
\textbf{III} & $%
\begin{array}{c}
\text{ } \\ 
\text{Existence} \\ 
\text{ }%
\end{array}%
$ & --------- & $\Pr \left( T_{\psi }\leq T_{\phi }\right) >\frac{1}{2}$ & $%
\Pr \left( T_{\psi }\leq T_{\phi }\right) <\frac{1}{2}$ \\ \hline
\textbf{III'} & $%
\begin{array}{c}
\text{ } \\ 
\text{Existence} \\ 
\text{ }%
\end{array}%
$ & --------- & $\delta >2\delta _{\min }$ & $\delta <2\delta _{\min }$ \\ 
\hline
\textbf{IV} & $%
\begin{array}{c}
\text{ } \\ 
\text{Fast} \\ 
\text{ }%
\end{array}%
$ & $0<\tau <\tau _{\ast }$ & $H\left( 0\right) >\frac{\delta _{\exp }}{%
\delta }$ & $H\left( 0\right) <\frac{\delta _{\exp }}{\delta }$ \\ \hline
\textbf{V} & $%
\begin{array}{c}
\text{ } \\ 
\text{Slow} \\ 
\text{ }%
\end{array}%
$ & $\tau ^{\ast }<\tau <\infty $ & $H\left( \infty \right) <\frac{\delta
_{\exp }}{\delta }$ & $H\left( \infty \right) >\frac{\delta _{\exp }}{\delta 
}$ \\ \hline
\end{tabular}
\end{center}

\bigskip

\textbf{Table 3}: Five pairs of diversity criteria regarding the sharp-restart algorithm. The Table's columns specify the features of each pair of criteria: to which timer parameters $\tau $ the criteria apply; and when does the application of the sharp-restart algorithm decrease or increase the diversity (of the output, with respect to that of the input). Row \textbf{I} -- criteria for general timers (section \ref{3}), where: $\bar{\Psi}(t)$ and $\bar{\Phi}(t)$ are the survival functions of two statistical distributions that are induced by the input's statistical distribution; for an integer Renyi exponent ($\epsilon=2,3, \cdots$), the first induced statistical distribution corresponds to the coincidence of $\epsilon$ IID copies of the input (see Eq. (\ref{303}), and the second induced statistical distribution corresponds to the minimum of $\epsilon$ IID copies of the input (see Eq. (\ref{304}). Row \textbf{II} -- existence criteria (section \ref{4}), where: $\mu_{\psi}$ and $\mu_{\phi}$ are, respectively, the means of the two induced statistical distributions. Row \textbf{III} -- existence criteria (section \ref{4}), where: $T_{\psi }$ and $T_{\phi }$ are mutually independent random variables that are drawn, respectively, from the two induced statistical distributions. Row \textbf{III'} -- an alternative representation of the existence criteria of row \textbf{III}, where: $\delta$ is the input's diversity, and $\delta_{\min}$ is the diversity of the minimum of two IID copies of the input.  Row \textbf{IV} -- criteria for fast timers (sections \ref{5} and \ref{6}), where: $H(0)$ is the limit value, at zero, of the input's hazard function; $\delta_{\exp}$ is the diversity of the unit-scale Exponential distribution (see Eq. (\ref{502})); and $\tau_{*}$ is the upper bound of the range of fast timers (see Eq. (\ref{601})). Row \textbf{V} -- criteria for slow timers (sections \ref{5} and \ref{6}), where: $H(\infty)$ is the limit value, at infinity, of the input's hazard function; and $\tau^{*}$ is the lower bound of the range of slow timers (see Eq. (\ref{602})).
\newpage

The various criteria display a trade-off between the information they require and the information they provide. For example, the first pair of criteria in Table 3 determine -- for any given timer $\tau $ -- if sharp restart decreases/increases diversity. However, this precise information comes with a high `price tag': full knowledge of the two induced statistical distributions (that are governed by the density functions $\psi \left( t\right)$ and $\phi \left( t\right)$). On the other hand, the second pair of criteria in Table 3 determine the very existence of timers with which sharp restart decreases/increases diversity -- doing so without pinpointing any particular such timer. This less detailed information comes at a far lower `price tag': knowing only the means of the two induced statistical distributions.

In the introduction to the first part of this duo \cite{EntSR}, it was noted that -- given a statistical distribution of interest -- scientists commonly strive to `separate the wheat from the chaff' by distilling the critical information conveyed by the distribution. This goal is attained by analyzing three principal aspects of the given statistical distribution: its mean behavior; its inherent randomness; and its tail behavior -- the likelihood of small and large outliers, also known as `rare events'. 

For sharp restart with high-frequency and low-frequency resetting (fast and slow timers), these three aspects come together beautifully in the neat mean-diversity classifications of Tables 1 and 2, and of Figure \ref{phase-diagram1}. Indeed, the mean-diversity classifications are based on three numbers emanating from the input's statistical distribution: its mean $\mu$ (quantifying the mean behavior); its diversity $\delta$ (quantifying the inherent randomness); and the limits of its hazard function -- the hazard limit $H(0)$ which corresponds to small outliers, and the hazard limit $H(\infty)$ which corresponds to large outliers. These classifications provide deep understanding and insights -- which are first of their kind -- regarding the intricate interplay between restart, mean-behavior, and randomness.

A series of recent works established a set of universal results for the mean-performance of restart protocols: general rules that determine the effect of restart on \emph{mean} completion times. Yet, to date, similar results for the \emph{randomness} of completion times were lacking. Going from mean to randomness, this duo presented a comprehensive stochasticity analysis of sharp restart: determining how it affects the entropy and the diversity of completion times. We hope that researchers will follow the path set in this duo, and will further explore the effect of restart on randomness. Possible directions for future research include: Poissonian restart protocols, general restart protocols, and optimization when resetting comes at a cost.

\bigskip

\textbf{Acknowledgments}. Shlomi Reuveni acknowledges support from the Israel Science Foundation (grant No. 394/19). This project has received funding from the European Research Council (ERC) under the European Union’s Horizon 2020 research and innovation program (Grant agreement No. 947731).  The authors thank Shira Yovel for her help in producing the figures. \\

\newpage 

\section{Methods}

\subsection{Derivation of Eqs. (\protect\ref{301}) and (\protect\ref{302})}

Set $p=F\left( \tau \right) $ and $q=\bar{F}\left( \tau \right) $, and note that $p+q=1$. Using the periodic parameterization $t=\tau n+u$ ($n=0,1,2,\cdots $ and $0\leq u<\tau $) of the time axis, and using Eq. (\ref{211}), we have:
\begin{equation}
\left. 
\begin{array}{l}
\mathcal{C}_{\epsilon }\left[ T_{R}\right] =\int_{0}^{\infty}f_{R}\left(t\right) ^{\epsilon }dt \\ 
\\ 
=\sum_{n=0}^{\infty }\int_{\tau n}^{\tau n+\tau }f_{R}\left( t\right)^{\epsilon}dt=\sum_{n=0}^{\infty }\int_{0}^{\tau }f_{R}\left( \tau n+u\right) ^{\epsilon }du \\ 
\\ 
=\sum_{n=0}^{\infty }\int_{0}^{\tau }\left[ q^{n}f\left( u\right) \right]^{\epsilon }du=\sum_{n=0}^{\infty }\left[ \left( q^{n}\right) ^{\epsilon}\int_{0}^{\tau }f\left( u\right) ^{\epsilon }du\right] \\ 
\\ 
=\left[ \sum_{n=0}^{\infty }\left( q^{\epsilon }\right) ^{n}\right] \cdot \int_{0}^{\tau }f\left( u\right) ^{\epsilon}du.
\end{array}
\right.  \label{A101}
\end{equation}
Note that
\begin{equation}
\sum_{n=0}^{\infty }\left( q^{\epsilon }\right)^{n}=\frac{1}{1-q^{\epsilon }}=\frac{1}{1-\bar{F}\left( \tau \right) ^{\epsilon }}.  \label{A102}
\end{equation}
Substituting Eq. (\ref{A102}) into the bottom line of Eq. (\ref{A101}) yields Eq. (\ref{301}).

To derive Eq. (\ref{302}), we re-write Eq. (\ref{A101}) as follows:
\begin{equation}
\left. 
\begin{array}{l}
\mathcal{C}_{\epsilon }\left[ T_{R}\right] =\left[ \sum_{n=0}^{\infty}\left( q^{\epsilon }\right) ^{n}\left( p^{\epsilon }\right) \right] \cdot \int_{0}^{\tau}\frac{1}{p^{\epsilon }}f\left( u\right) ^{\epsilon }du \\ 
\\ 
=\left[ \sum_{n=0}^{\infty }\left( q^{n}p\right) ^{\epsilon }\right] \cdot \int_{0}^{\tau }\left[ \frac{1}{p}f\left(u\right) \right] ^{\epsilon }du.
\end{array}
\right.  \label{A103}
\end{equation}
Consider the random variable $N$ of Eq. (\ref{212}), whose probability distribution is $\Pr \left( N=n\right) =q^{n}p$ ($n=0,1,2,\cdots $); for this random variable note that
\begin{equation}
\mathcal{C}_{\epsilon }\left[ N\right] =\sum_{n=0}^{\infty }\Pr \left(N=n\right) ^{\epsilon }=\sum_{n=0}^{\infty }\left( q^{n}p\right) ^{\epsilon }. \label{A104}
\end{equation}
Consider the random variable $U$ of Eq. (\ref{212}) -- whose density function is $\frac{1}{p}f\left( u\right) $ ($0\leq u<\tau $); for this random variable note that
\begin{equation}
\mathcal{C}_{\epsilon }\left[ U\right] =\int_{0}^{\tau }\left[ \frac{1}{p} f\left( u\right) \right] ^{\epsilon }du.  \label{A105}
\end{equation}
Substituting Eqs. (\ref{A104}) and (\ref{A105}) into the bottom line of Eq. (\ref{A103}) yields
\begin{equation}
\mathcal{C}_{\epsilon }\left[ T_{R}\right] =\mathcal{C}_{\epsilon }\left[ N\right] \cdot \mathcal{C}_{\epsilon }\left[ U\right].  \label{A106}
\end{equation}
In turn, Eq. (\ref{A106}) yields Eq. (\ref{302}).

\subsection{Derivation of Eqs. (\protect\ref{403}) and (\protect\ref{404})}

Eq. (\ref{305}) implies that

\begin{equation}
\frac{C\left( \tau \right) -\gamma }{\gamma }=\frac{\Psi \left( \tau \right) -\Phi \left( \tau \right) }{\Phi \left( \tau \right) }.  \label{A111}
\end{equation}
Multiplying both sides of Eq. (\ref{A111}) by $\phi _{\max }\left( \tau \right) =2\Phi \left( \tau \right) \phi \left( \tau \right) $ further implies that
\begin{equation}
\frac{C\left( \tau \right) -\gamma }{\gamma }\phi _{\max }\left( \tau \right) =2\Psi \left( \tau \right) \phi \left( \tau \right) -\phi _{\max}\left( \tau \right) . \label{A112}
\end{equation}
In turn, integrating Eq. (\ref{A112}) over the positive half-line yields
\begin{equation}
\left. 
\begin{array}{l}
\int_{0}^{\infty }\left[ \frac{C\left( \tau \right) -\gamma }{\gamma }\right]
\phi _{\max }\left( \tau \right) d\tau \\ 
\\ 
=\int_{0}^{\infty }2\Psi \left( \tau \right) \phi \left( \tau \right) d\tau -\int_{0}^{\infty }\phi _{\max }\left( \tau \right) d\tau \\ 
\\ 
=2\int_{0}^{\infty }\Psi \left( \tau \right) \phi \left( \tau \right) d\tau
-1.
\end{array}
\right.  \label{A113}
\end{equation}
For independent random variables $T_{\psi }$ and $T_{\phi }$, using
conditioning, note that
\begin{equation}
\left. 
\begin{array}{l}
\Pr \left( T_{\psi }\leq T_{\phi}\right) =\int_{0}^{\infty }\Pr \left(T_{\psi }\leq \tau \vert T_{\phi }=\tau \right) \phi \left( \tau \right) d\tau \\ 
\\ 
=\int_{0}^{\infty }\Psi \left( \tau \right) \phi \left( \tau \right) d\tau.
\end{array}
\right.  \label{A114}
\end{equation}
Substituting Eq. (\ref{A114}) into the bottom line of Eq. (\ref{A113})
yields Eq. (\ref{403}).

Similarly to Eq. (\ref{A114}) note that
\begin{equation}
\left. 
\begin{array}{l}
\Pr \left( T_{\psi }\leq T_{\phi }\right) =\Pr \left( T_{\phi }>T_{\psi}\right)  \\ 
\\ 
=\int_{0}^{\infty }\Pr \left( T_{\phi }>t \vert T_{\psi}=t \right) \psi \left( t\right) dt \\ 
\\ 
=\int_{0}^{\infty }\bar{\Phi}\left( t \right) \psi \left( t\right) dt=\int_{0}^{\infty }\left[ \bar{F}\left( t\right) ^{\epsilon }\right] \left[\frac{1}{\gamma }f\left( t\right) ^{\epsilon }\right] dt \\ 
\\ 
=\frac{1}{2^{\epsilon }\gamma }\int_{0}^{\infty }\left[ 2\bar{F}\left(t\right) f\left( t\right) \right] ^{\epsilon }dt;
\end{array}
\right.   \label{A115}
\end{equation}
in the third line of Eq. (\ref{A115}) we used Eqs. (\ref{303}) and (\ref{304}). The term $2\bar{F}\left( t\right) f\left( t\right) $ -- which appears in the bottom line of Eq. (\ref{A115}) -- is the density function of the random variable $\min \left\{T_{1},T_{2}\right\} $, the minimum of two IID copies of the input $T$. Hence, Eq. (\ref{A115}) implies that
\begin{equation}
\Pr \left( T_{\psi }\leq T_{\phi }\right) =\frac{1}{2^{\epsilon }}\frac{\mathcal{C}_{\epsilon }\left[ \min \left\{ T_{1},T_{2}\right\} \right] }{\mathcal{C}_{\epsilon }\left[ T\right] }. \label{A117}
\end{equation}
In turn, Eq. (\ref{A117}) implies that
\begin{equation}
2\Pr \left( T_{\psi }\leq T_{\phi }\right) =\left( \frac{\mathcal{D}_{\epsilon }\left[ T\right] }{2\mathcal{D}_{\epsilon }\left[ \min \left\{T_{1},T_{2}\right\} \right] }\right) ^{\epsilon -1}.  \label{A118}
\end{equation}
Substituting Eq. (\ref{A118}) into Eq. (\ref{403}) -- while using the shorthand notations $\delta =\mathcal{D}_{\epsilon }\left[ T\right] $ and $\delta _{\min }=\mathcal{D}_{\epsilon }\left[ \min \left\{T_{1},T_{2}\right\} \right] $ -- yields Eq. (\ref{404}).

\subsection{Derivation of Eqs. (\protect\ref{503}) and (\protect\ref{504})}

Using the density function $\psi \left( t\right) =\frac{1}{\gamma }f\left(
t\right) ^{\epsilon }$ of Eq. (\ref{303}), the density function $\phi \left(
t\right) =\epsilon \bar{F}\left( t\right) ^{\epsilon -1}f\left( t\right) $
induced by the survival function of Eq. (\ref{304}), the input's hazard
function of Eq. (\ref{501}), Eq. (\ref{502}), and the connection between the
coincidence likelihood and the diversity (recall Eq. (\ref{222})), note that:%
\begin{equation}
\left. 
\begin{array}{l}
\frac{\psi \left( t\right) }{\phi \left( t\right) }=\frac{\frac{1}{\gamma }%
f\left( t\right) ^{\epsilon }}{\epsilon \bar{F}\left( t\right) ^{\epsilon
-1}f\left( t\right) } \\ 
\\ 
=\frac{1}{\gamma \epsilon }\left[ \frac{f\left( t\right) }{\bar{F}\left(
t\right) }\right] ^{\epsilon -1}=\frac{1}{\gamma \epsilon }H\left( t\right)
^{\epsilon -1} \\ 
\\ 
=\frac{1}{\gamma }\left[ \frac{1}{\delta_{\exp} }H\left( t\right) \right] ^{\epsilon
-1}=\left[ \frac{\delta }{\delta_{\exp} }H\left( t\right) \right] ^{\epsilon -1}.
\end{array}
\right.  \label{A120}
\end{equation}

Taking the limit $\tau \rightarrow 0$ in the middle part of Eq. (\ref{305}),
while using L'Hospital's rule and Eq. (\ref{A120}), we have:%
\begin{equation}
\left. 
\begin{array}{l}
\lim_{\tau \rightarrow 0}\frac{C\left( \tau \right) }{\gamma }=\lim_{\tau
\rightarrow 0}\frac{\Psi \left( \tau \right) }{\Phi \left( \tau \right) } \\ 
\\ 
=\lim_{t\rightarrow 0}\frac{\psi \left( t\right) }{\phi \left( t\right) }%
=\lim_{t\rightarrow 0}\frac{1}{\gamma }\left[ \frac{1}{\delta_{\exp} }H\left( t\right) %
\right] ^{\epsilon -1} \\ 
\\ 
=\frac{1}{\gamma }\left[ \frac{1}{\delta_{\exp} }\lim_{t\rightarrow 0}H\left( t\right) 
\right] ^{\epsilon -1}.
\end{array}
\right.  \label{A121}
\end{equation}
Eq. (\ref{A121}) implies that%
\begin{equation}
C\left( 0\right) =\left[ \frac{1}{\delta_{\exp} }H\left( 0\right) \right] ^{\epsilon
-1}.  \label{A122}
\end{equation}%
In turn, using the connection between the coincidence likelihood and the
diversity (recall Eq. (\ref{222})), Eq. (\ref{A122}) implies Eq. (\ref{503}). 

Taking the limit $\tau \rightarrow \infty $ in the middle part of Eq. (\ref%
{305}) yields%
\begin{equation}
\lim_{\tau \rightarrow \infty }\frac{C\left( \tau \right) }{\gamma }%
=\lim_{\tau \rightarrow \infty }\frac{\Psi \left( \tau \right) }{\Phi \left(
\tau \right) }=1,  \label{A123}
\end{equation}%
and hence $C\left( \infty \right) =\gamma $. Eq. (\ref{305}) implies that%
\begin{equation}
\frac{C\left( \tau \right) -\gamma }{\gamma }=\frac{\bar{\Phi}\left( \tau
\right) -\bar{\Psi}\left( \tau \right) }{\Phi \left( \tau \right) }.
\label{A124}
\end{equation}%
In turn, Eq. (\ref{A124}) implies that%
\begin{equation}
\frac{C\left( \tau \right) -\gamma }{\gamma \bar{\Phi}\left( \tau \right) }=%
\frac{1}{\Phi \left( \tau \right) }\left[ 1-\frac{\bar{\Psi}\left( \tau
\right) }{\bar{\Phi}\left( \tau \right) }\right] .  \label{A125}
\end{equation}%
Taking the limit $\tau \rightarrow \infty $ in Eq. (\ref{A125}), while using
L'Hospital's rule and Eq. (\ref{A120}), we have:%
\begin{equation}
\left. 
\begin{array}{l}
\lim_{\tau \rightarrow \infty }\frac{C\left( \tau \right) -\gamma }{\gamma 
\bar{\Phi}\left( \tau \right) }=1-\lim_{\tau \rightarrow \infty }\frac{\bar{%
\Psi}\left( \tau \right) }{\bar{\Phi}\left( \tau \right) } \\ 
\\ 
=1-\lim_{t\rightarrow \infty }\frac{\psi \left( t\right) }{\phi \left(
t\right) }=1-\lim_{t\rightarrow \infty }\frac{1}{\gamma }\left[ \frac{1}{\delta_{\exp} 
}H\left( t\right) \right] ^{\epsilon -1} \\ 
\\ 
=1-\frac{1}{\gamma }\left[ \frac{1}{\delta_{\exp} }\lim_{t\rightarrow \infty }H\left(
t\right) \right] ^{\epsilon -1}.
\end{array}
\right.  \label{A126}
\end{equation}
Eq. (\ref{A126}) implies that
\begin{equation}
\lim_{\tau \rightarrow \infty }\frac{C\left( \tau \right) -\gamma }{\bar{\Phi
}\left( \tau \right) }=\gamma -\left[ \frac{1}{\delta_{\exp} }H\left( \infty \right) 
\right] ^{\epsilon -1}.  \label{A127}
\end{equation}
In turn, using the connection between the coincidence likelihood and the
diversity (recall Eq. (\ref{222})), Eq. (\ref{A127}) implies Eq. (\ref{504}).

\subsection{Derivation of Eqs. (\protect\ref{603}) and (\protect\ref{604})}

Eq. (\ref{A120}), Eq. (\ref{502}), and the connection between the
coincidence likelihood and the diversity (recall Eq. (\ref{222})) imply
that: 
\begin{equation}
\left. 
\begin{array}{l}
\Psi \left( \tau \right) -\Phi \left( \tau \right) =\int_{0}^{\tau }\psi
\left( t\right) dt-\int_{0}^{\tau }\phi \left( t\right) dt \\ 
\\ 
=\int_{0}^{\tau }\left[ \psi \left( t\right) -\phi \left( t\right) \right]
dt=\int_{0}^{\tau }\left[ \frac{\psi \left( t\right) }{\phi \left( t\right) }%
-1\right] \phi \left( t\right) dt \\ 
\\ 
=\int_{0}^{\tau }\left\{ \left[ \frac{\delta }{\delta_{\exp} }H\left( t\right) \right]
^{\epsilon -1}-1\right\} \phi \left( t\right) dt \\ 
\\ 
=\left( \frac{\delta }{\delta_{\exp} }\right) ^{\epsilon -1}\int_{0}^{\tau }\left[
H\left( t\right) ^{\epsilon -1}-\left( \frac{\delta_{\exp} }{\delta }\right)
^{\epsilon -1}\right] \phi \left( t\right) dt \\ 
\\ 
=\frac{1}{\gamma \epsilon }\int_{0}^{\tau }\left[ H\left( t\right)
^{\epsilon -1}-\left( \frac{\delta_{\exp} }{\delta }\right) ^{\epsilon -1}\right] \phi
\left( t\right) dt.
\end{array}
\right.  \label{A141}
\end{equation}
Substituting Eq. (\ref{A141}) into Eq. (\ref{A111}) yields Eq. (\ref{603}).

Eq. (\ref{A120}), Eq. (\ref{502}), and the connection between the
coincidence likelihood and the diversity (recall Eq. (\ref{222})) imply
that: 
\begin{equation}
\left. 
\begin{array}{l}
\bar{\Phi}\left( \tau \right) -\bar{\Psi}\left( \tau \right) =\int_{\tau
}^{\infty }\phi \left( t\right) dt-\int_{\tau }^{\infty }\psi \left(
t\right) dt \\ 
\\ 
=\int_{\tau }^{\infty }\left[ \phi \left( t\right) -\psi \left( t\right) %
\right] dt=\int_{\tau }^{\infty }\left[ 1-\frac{\psi \left( t\right) }{\phi
\left( t\right) }\right] \phi \left( t\right) dt \\ 
\\ 
=\int_{\tau }^{\infty }\left\{ 1-\left[ \frac{\delta }{\delta_{\exp} }H\left( t\right) %
\right] ^{\epsilon -1}\right\} \phi \left( t\right) dt \\ 
\\ 
=\left( \frac{\delta }{\delta_{\exp} }\right) ^{\epsilon -1}\int_{\tau }^{\infty }%
\left[ \left( \frac{\delta_{\exp} }{\delta }\right) ^{\epsilon -1}-H\left( t\right)
^{\epsilon -1}\right] \phi \left( t\right) dt \\ 
\\ 
=\frac{1}{\gamma \epsilon }\int_{\tau }^{\infty }\left[ \left( \frac{\delta_{\exp} }{%
\delta }\right) ^{\epsilon -1}-H\left( t\right) ^{\epsilon -1}\right] \phi
\left( t\right) dt.
\end{array}%
\right.   \label{A142}
\end{equation}%
Substituting Eq. (\ref{A142}) into Eq. (\ref{A124}) yields Eq. (\ref{604}).


\newpage

\begin{thebibliography}{99}

\bibitem{EntSR} Eliazar, Iddo, and Shlomi Reuveni. Entropy of Sharp Restart. arXiv preprint arXiv:2207.02085 (2022).

\bibitem{Pal2017}Pal, Arnab, and Shlomi Reuveni. First passage under restart. Physical review letters 118, no. 3 (2017): 030603.

\bibitem{Che2018}Chechkin, A., and IM Sokolov. Random search with resetting: a unified renewal approach. Physical review letters 121, no. 5 (2018): 050601.

\bibitem{Eva2020}Evans, Martin R., Satya N. Majumdar, and Grégory Schehr. Stochastic resetting and applications. Journal of Physics A: Mathematical and Theoretical 53, no. 19 (2020): 193001.

\bibitem{Kus2014}Kuśmierz, Łukasz, Satya N. Majumdar, Sanjib Sabhapandit, and Grégory Schehr. First order transition for the optimal search time of Lévy flights with resetting. Physical review letters 113, no. 22 (2014): 220602.

\bibitem{Kus2015}Kuśmierz, Łukasz, and Ewa Gudowska-Nowak. Optimal first-arrival times in Lévy flights with resetting. Physical Review E 92, no. 5 (2015): 052127.

\bibitem{Rot2015}Rotbart, Tal, Shlomi Reuveni, and Michael Urbakh. Michaelis-Menten reaction scheme as a unified approach towards the optimal restart problem. Physical Review E 92, no. 6 (2015): 060101.

\bibitem{Bha2016}Bhat, Uttam, Caterina De Bacco, and S. Redner. Stochastic search with Poisson and deterministic resetting. Journal of Statistical Mechanics: Theory and Experiment 2016, no. 8 (2016): 083401.

\bibitem{Bes2020}Besga, Benjamin, Alfred Bovon, Artyom Petrosyan, Satya N. Majumdar, and Sergio Ciliberto. Optimal mean first-passage time for a Brownian searcher subjected to resetting: experimental and theoretical results. Physical Review Research 2, no. 3 (2020): 032029.

\bibitem{TalF2020}Tal-Friedman, Ofir, Arnab Pal, Amandeep Sekhon, Shlomi Reuveni, and Yael Roichman. Experimental realization of diffusion with stochastic resetting. The journal of physical chemistry letters 11, no. 17 (2020): 7350-7355.

\bibitem{Ray_2020} Ray, S. and Reuveni, S., 2020. Diffusion with resetting in a logarithmic potential. J. Chem. Phys. 152, 234110.

\bibitem{Che2021} Chelminiak, Przemyslaw. "Non-linear diffusion with stochastic resetting." arXiv preprint arXiv:2107.14680 (2021).

\bibitem{Can2021}Cantisán, Julia, Jesús M. Seoane, and Miguel AF Sanjuán. Stochastic resetting in the Kramers problem: a Monte Carlo approach. Chaos, Solitons $\&$ Fractals 152 (2021): 111342.

\bibitem{Bon2021}Bonomo, Ofek Lauber, Arnab Pal, and Shlomi Reuveni. Mitigating long queues and waiting times with service resetting. arXiv preprint arXiv:2111.02097 (2021).

\bibitem{Che2022}Chen, Hanshuang, and Feng Huang. First passage of a diffusing particle under stochastic resetting in bounded domains with spherical symmetry. Physical Review E 105, no. 3 (2022): 034109.

\bibitem{Rad2022}Radice, Mattia. Diffusion processes with Gamma-distributed resetting and non-instantaneous returns. Journal of Physics A: Mathematical and Theoretical 55, no. 22 (2022): 224002.

\bibitem{Yin2022}Yin, Ruoyu, and Eli Barkai. Restart expedites quantum walk hitting times. arXiv preprint arXiv:2205.01974 (2022).


\bibitem{MP1SR} Eliazar, Iddo, and Shlomi Reuveni. Mean-performance of sharp restart I: Statistical roadmap. Journal of Physics A: Mathematical and Theoretical 53, no. 40 (2020): 405004.

\bibitem{MP2SR} Eliazar, Iddo, and Shlomi Reuveni. Mean-performance of sharp restart: II. Inequality roadmap. Journal of Physics A: Mathematical and Theoretical 54, no. 35 (2021): 355001.

\bibitem{TBSR} Eliazar, Iddo, and Shlomi Reuveni. Tail-behavior roadmap for sharp restart. Journal of Physics A: Mathematical and Theoretical 54, no. 12 (2021): 125001.


\bibitem{Five} Eliazar, Iddo. Five degrees of randomness. Physica A: Statistical Mechanics and its Applications 568 (2021): 125662.


\bibitem{Sim} Simpson, Edward H. Measurement of diversity. Nature 163, no. 4148 (1949): 688-688.

\bibitem{Hir1} Hirschman, Albert O. National power and the structure of foreign trade. University of Califorina Press, Berkeley, 1945.

\bibitem{BDH} Bell, R. J., P. Dean, and D. C. Hibbins-Butler. Localization of normal modes in vitreous silica, germania and beryllium fluoride. Journal of Physics C: Solid State Physics 3, no. 10 (1970): 2111.

\bibitem{BD} Bell, R. J., and P. Dean. The structure of vitreous silica: Validity of the random network theory. Philosophical Magazine 25, no. 6 (1972): 1381-1398.

\bibitem{BPP} Bosyk, Gustavo Martín, M. Portesi, and A. Plastino. Collision entropy and optimal uncertainty. Physical Review A 85, no. 1 (2012): 012108.


\bibitem{Hil} Hill, Mark O. Diversity and evenness: a unifying notation and its consequences. Ecology 54, no. 2 (1973): 427-432.

\bibitem{Pee} Peet, Robert K. The measurement of species diversity. Annual review of ecology and systematics 5, no. 1 (1974): 285-307.

\bibitem{Mag} Magurran, Anne E. Ecological diversity and its measurement. Princeton university press, 1988.

\bibitem{Jos} Jost, L., 2006. Entropy and diversity. Oikos 113, no. 2, pp.363-375.

\bibitem{LL} Legendre, Pierre, and Louis Legendre. Numerical ecology. Elsevier, 2012.


\bibitem{Ren} Renyi, Alfred. On measures of information and entropy. In Proceedings of the 4th Berkeley symposium on mathematics, statistics and probability, pp. 547-561, vol. 1, no. 547. 1961.

\bibitem{Len} Lenzi, E. K., R. S. Mendes, and L. R. Da Silva. Statistical mechanics based on Renyi entropy. Physica A: Statistical Mechanics and its Applications 280, no. 3-4 (2000): 337-345.

\bibitem{Zyc} Zyczkowski, Karol. Renyi extrapolation of Shannon entropy. Open Systems and Information Dynamics 10, no. 03 (2003): 297-310.


\bibitem{Bol} L. Boltzmann, Vorlesungen uber gastheorie (Lectures on gas theory). J.A. Barth Berlin: part I 1896; part II 1898.

\bibitem{Gib} J.W. Gibbs, Elementary principles in statistical mechanics. Yale University Press, 1902.

\bibitem{Sha} Shannon, Claude Elwood. A mathematical theory of communication. The Bell system technical journal 27, no. 3 (1948): 379-423.


\bibitem{Slow_Time_1}Reuveni, S., Urbakh, M. and Klafter, J., 2014. Role of substrate unbinding in Michaelis-Menten enzymatic reactions. Proceedings of the National Academy of Sciences, 111(12), pp.4391-4396.

\bibitem{Slow_Time_2}Pal, Arnab, and V. V. Prasad. Landau-like expansion for phase transitions in stochastic resetting. Physical Review Research 1, no. 3 (2019): 032001.

\bibitem{Slow_Time_3} Ray, S., Mondal, D. and Reuveni, S., 2019. Péclet number governs transition to acceleratory restart in drift-diffusion. Journal of Physics A: Mathematical and Theoretical, 52(25), p.255002.

\bibitem{Slow_Time_4}Ahmad, S., Nayak, I., Bansal, A., Nandi, A. and Das, D., 2019. First passage of a particle in a potential under stochastic resetting: A vanishing transition of optimal resetting rate. Physical Review E, 99(2), p.022130.

\bibitem{Slow_Time_5}Pal, Arnab, Iddo Eliazar, and Shlomi Reuveni. First passage under restart with branching. Physical review letters 122, no. 2 (2019): 020602.

\bibitem{Slow_Time_6}Ray, Somrita, Debasish Mondal, and Shlomi Reuveni. Péclet number governs transition to acceleratory restart in drift-diffusion. Journal of Physics A: Mathematical and Theoretical 52, no. 25 (2019): 255002.

\bibitem{Slow_Time_7}Bressloff, Paul C. Queueing theory of search processes with stochastic resetting. Physical Review E 102, no. 3 (2020): 032109.

\bibitem{Slow_Time_8}Singh, R. K., R. Metzler, and T. Sandev. Resetting dynamics in a confining potential. Journal of Physics A: Mathematical and Theoretical 53, no. 50 (2020): 505003.

\bibitem{Slow_Time_9}Ray, Somrita, and Shlomi Reuveni. Resetting transition is governed by an interplay between thermal and potential energy. The Journal of Chemical Physics 154, no. 17 (2021): 171103.

\bibitem{Slow_Time_10}Méndez, Vicenç, Axel Masó-Puigdellosas, and Daniel Campos. Nonstandard diffusion under Markovian resetting in bounded domains. Physical Review E 105, no. 5 (2022): 054118.

\bibitem{Slow_Time_11}Pal, A., Kostinski, S. and Reuveni, S., 2022. The inspection paradox in stochastic resetting. Journal of Physics A: Mathematical and Theoretical, 55(2), p.021001.

\bibitem{Hir2} Hirschman, A. O. The Paternity of an Index. The American Economic Review 54, no. 5 (1964): 761-762.

\bibitem{MAAD} Eliazar, Iddo. From moving averages to anomalous diffusion: a Renyi-entropy approach. Journal of Physics A: Mathematical and Theoretical 48, no. 3 (2014): 03FT01.

\bibitem{JMB} Jelinek, Fred, Robert L. Mercer, Lalit R. Bahl, and James K. Baker. Perplexity—a measure of the difficulty of speech recognition tasks. The Journal of the Acoustical Society of America 62, no. S1 (1977): S63-S63.

\bibitem{Par} Pareto, Vilfredo. Cours d'économie politique. Vol. 1. Librairie Droz, 1964. Pareto, Vilfredo. Manual of political economy: a critical and variorum edition. Oxford University Press, 2014.

\bibitem{New} Newman, Mark EJ. Power laws, Pareto distributions and Zipf's law. Contemporary physics 46, no. 5 (2005): 323-351.

\bibitem{CSN} Clauset, Aaron, Cosma Rohilla Shalizi, and Mark EJ Newman. Power-law distributions in empirical data. SIAM review 51, no. 4 (2009): 661-703.

\bibitem{Har} Hardy, Michael. Pareto’s law. The Mathematical Intelligencer 32, no. 3 (2010): 38-43.

\bibitem{Arn} Arnold, Barry. Pareto distributions. CRC Press, 2015.


\bibitem{Wei} Weibull, Waloddi. A statistical distribution function of wide applicability. Journal of applied mechanics 103, no. 107 (1951): 293-297.

\bibitem{FT} Fisher, Ronald Aylmer, and Leonard Henry Caleb Tippett. Limiting forms of the frequency distribution of the largest or smallest member of a sample. In Mathematical proceedings of the Cambridge philosophical society, vol. 24, no. 2, pp. 180-190. Cambridge University Press, 1928.

\bibitem{Gne} Gnedenko, B. V. On limit theorems for a random number of random variables. In Probability Theory and Mathematical Statistics, pp. 167-176. Springer, Berlin, Heidelberg, 1983.

\bibitem{Gal} Galambos, Janos. The asymptotic theory of extreme order statistics. Krieger, Melbourne, Florida, 1987.

\bibitem{BGS} Beirlant, Jan, Yuri Goegebeur, Johan Segers, and Jozef L. Teugels. Statistics of extremes: theory and applications. Vol. 558. John Wiley \& Sons, 2004.

\bibitem{RT} Reiss, Rolf-Dieter, Michael Thomas, and R. D. Reiss. Statistical analysis of extreme values. Birkhauser, Basel, 2007.

\bibitem{MXJ} Murthy, DN Prabhakar, Min Xie, and Renyan Jiang. Weibull models. Vol. 505. John Wiley \& Sons, 2004.

\bibitem{Rin} Rinne, Horst. The Weibull distribution: a handbook. Chapman and Hall/CRC, 2008.

\bibitem{McC} McCool, John I. Using the Weibull distribution: reliability, modeling, and inference. Vol. 950. John Wiley \& Sons, 2012.


\bibitem{WW} Williams, Graham, and David C. Watts. Non-symmetrical dielectric relaxation behaviour arising from a simple empirical decay function. Transactions of the Faraday society 66 (1970): 80-85.

\bibitem{Phi} Phillips, J. C. Stretched exponential relaxation in molecular and electronic glasses. Reports on Progress in Physics 59, no. 9 (1996): 1133.

\bibitem{CK} Kalmykov, Yuri and William T. Coffey (Eds.). Fractals, Diffusion, and Relaxation in Disordered Complex Systems. John Wiley \& Sons, 2006.

\bibitem{KP} Kalbfleisch, John D., and Ross L. Prentice. The statistical analysis of failure time data. John Wiley \& Sons, 2011.

\bibitem{KK} Kleinbaum, David G., and Mitchel Klein. Survival analysis: a self-learning text. Springer, 2012.

\bibitem{Col} Collett, David. Modelling survival data in medical research. CRC press, 2015.

\bibitem{BP} Barlow, Richard E., and Frank Proschan. Mathematical theory of reliability. Society for Industrial and Applied Mathematics, 1996.

\bibitem{Fin} Finkelstein, Maxim. Failure rate modelling for reliability and risk. Springer Science \& Business Media, 2008.

\bibitem{Dhi} Dhillon, Balbir S. Engineering systems reliability, safety, and maintenance: an integrated approach. CRC Press, 2017. 

\end{thebibliography}
\end{document}